\documentclass[preprint,showpacs,12pt,nofootinbib]{revtex4}

\usepackage{amsmath,amssymb,amsfonts}
\usepackage{graphicx}% Include figure files
\usepackage{hhline}
\usepackage{bm}
\usepackage{amsmath}
\usepackage{epsfig}
\usepackage{graphicx}

\usepackage{tabularx}
\usepackage{lscape}
\usepackage{rotating}
\usepackage{pstricks}

\usepackage{footnote}
\usepackage{hyperref}

\usepackage{pdfpages}
\usepackage{color} 

\usepackage{enumitem}

\newcommand{\bfvec}[1]{\hbox{\boldmath$#1$\unboldmath}}

\newcommand{\QCD}{\mathrm{QCD}}
\newcommand{\eff}{\mathrm{eff}}

\begin{document}
\title{From QCD phenomenology to nuclear physics phenomenology: the chiral confining model}

\author{Guy Chanfray$^{1}$, Magda Ericson$^{1,2}$, Hubert Hansen$^{1}$, J\'er\^ome Margueron$^{3}$ and Marco Martini$^{4,5}$}

\affiliation{$^{1}$ University Claude Bernard Lyon 1, CNRS/IN2P3, IP2I Lyon, UMR 5822, 69622  Villeurbanne, France \\
$^{2}$ Theory Division, CERN, CH-12111 Geneva, Switzerland\\
$^{3}$ International Research Laboratory on Nuclear Physics and Astrophysics,
Michigan State University and CNRS, East Lansing, MI 48824, USA\\
$^{4}$ IPSA-DRII,  63 boulevard de Brandebourg, 94200 Ivry-sur-Seine, France \\
$^{5}$ Sorbonne Universit\'e, CNRS/IN2P3, Laboratoire
de Physique Nucl\'eaire et de Hautes Energies (LPNHE), 75005 Paris, France} 

\begin{abstract}
We present a theoretical framework allowing to make an explicit connection between the phenomenology of QCD, namely the properties of the gluon correlator and Wilson loops, and a particular relativistic model for the description of nuclear matter and neutron stars, the chiral confining model. Starting with the Field Correlator Method, which incorporates explicitly and simultaneously confinement and chiral symmetry breaking, we describe how to obtain the response of the  composite nucleon to the nuclear scalar field, the relative role of confinement and chiral symmetry breaking in the in-medium nucleon mass evolution, generating the three-body forces needed for the saturation mechanism.

\end{abstract}

\maketitle
The aim of this paper is to discuss the microscopic foundations of a particular model devoted to the study of nuclear and neutron star physics, which incorporates at a phenomenological level the main non-perturbative properties of the underlying Quantum Chromodynamics (QCD), namely confinement and chiral symmetry breaking (CSB). The central motivation is to relate the main parameters governing the saturation mechanism and the stabilization of nuclear matter (three-body forces) arising from the evolution with density of the in-medium nucleon mass and the chirally broken vacuum (i.e., the scalar effective potential), to the non perturbative gluon correlator. Such an approach can be called "fundamental phenomenology" since the main dynamical input is provided by the properties of the lowest order (gaussian) gluon correlator obtained from lattice data measurement, without any reference to a particular non-perturbative microscopic model of the QCD vacuum or ad hoc confining model of the nucleon. The main phenomenological QCD  input are thus the string tension $\sigma$ and the gluon condensate $\mathcal{G}_2$  or alternatively, the string tension $\sigma$ and the string radius $T_g$. 

The first section gives a summary of the phenomenological nuclear physics model under consideration, namely the \textit{chiral confining model} . In the second section we present the Field Correlator Method (FCM) allowing the construction of an effective quark hamiltonian and its connection with Wilson loop properties. We also present a prescription which allows to disentangle the string-like confining piece of the interaction, $V_C(R) \sim \sigma R $ ($R$ is the length of the string confining the quarks inside the bound nucleon) from its self-interaction piece, $V_{CSB}(r)$, generating  a BCS-like broken vacuum, i.e., the vacuum of constituent quarks with mass $M\sim \sigma T_g$. In the third section we give a general discussion of the existence of a non trivial solution  of the gap equation. Although the previous non-perturbative self-interaction potential, whose strength behaves as $4\pi\sigma T_g^4$, is the main contributor, the additional contribution of the perturbative gluon exchange is required to obtain a quantitatively acceptable solution. For practical nuclear calculations we approximate (after Fierz transformation) the previously CSB potential generating chiral symmetry breaking and condensation of $q\bar{q}$ pairs by an equivalent NJL model, which is discussed in details in the fourth section. In the fifth section we derive the bound-state equation for a constituent quark moving in the BCS-like vacuum and subjected to the confining potential $V_C (R)$ with origin at the string junction of the Y-shaped nucleon. We thus derive the density evolution of the nucleon mass and the determination of the response parameters of the nucleon to  the nuclear scalar field (scalar coupling constant and susceptibility) as well as  the various parameters ($g_A$, $g_{\pi NN}$, form factor cutoffs,...) entering the QCD-connected version of the \textsl{chiral confining model}. In the sixth section we discuss the specific contribution of the pionic self-energy to the nucleon mass and the related delicate questions of the effects of the finite size of the $q\bar{q}$ pion and of the center of mass corrections.

%\section{Motivation for the \textsl{chiral confining model}}
\section{Introduction: motivation and status of the phenomenological chiral confining model}

In the early 2000's some of the authors of the present paper (G.C and M.E) developed 
an approach that aimed to incorporate the main non-perturbative properties of QCD, namely chiral symmetry breaking and color confinement, in the description of nuclear matter and ultimately finite nuclei and neutron stars. This approach, that we now call the \textsl{chiral confining model}, is based on three physical pillars:
\begin{enumerate}[label=(\roman*)]
\item The scalar field $s$, associated with the radial fluctuation of the chiral condensate, is identified with the sigma meson of relativistic theories as originally proposed in Ref.~\cite{Chanfray2001}. This proposal, which gives a plausible answer to the long standing problem of the chiral status of the sigma meson, $\sigma_W$, in relativistic mean-field approaches initiated by Walecka and collaborators~\cite{SerotWalecka1986,Walecka1997}, has also the merit of respecting all the desired chiral constraints. In particular the correspondence $s\equiv \sigma_W$, generates a coupling of the scalar field to the derivatives of the pion field. Hence the radial mode decouples from low-energy pions (as the pion is a quasi-Goldstone boson) whose dynamics is governed by chiral perturbation theory. A detailed discussion of this somewhat subtle topic is given in Ref.~\cite{Martini2006}.
\item The quark substructure of the nucleon is reflected by its polarizability,  in presence of the nuclear scalar field generating a repulsive three-nucleon force providing an efficient saturation mechanism \cite{Chanfray2005,Chanfray2007,Ericson2007,Massot2008,Massot2009,Massot2012, Rahul,Cham1,Cham2,Universe,Chanfray2024}. The physical motivation is rather obvious since, in reality, the composite nature of nucleon  reacts to the nuclear environment by readjusting its confined quark substructure, as pointed out in the pioneering paper of P. Guichon~\cite{Guichon1988} (this paper was also at the origin of the Quark Meson Coupling (QMC) model for nuclear matter, which has been successfully applied to finite nuclei by mapping it to Skyrme energy density functional~\cite{Guichon2004,Guichon2006,Stone}). There is indeed an absolute necessity to incorporate this effect to eliminate a well identified problem concerning the nuclear saturation with usual chiral effective theories~\cite{Boguta83,KM74,BT01,C03}. Independently of the particular chiral model, in the nuclear medium one can move away from the minimum of the vacuum effective potential (Mexican hat potential), i.e., into a region of smaller curvature. This single effect equivalent to the lowering of the sigma mass, destroys the stability, creating problems for the applicability of such effective theories in the nuclear context. The effect can be associated to a $s^3$ tadpole diagram generating attractive three-body forces destroying saturation even if the repulsive three-body force from the Walecka mechanism is present.
\item The associated response parameters, namely the nucleon scalar coupling constant, $g_S$, and the nucleonic scalar susceptibility, $\kappa_{NS}$, can be related to two chiral properties of the nucleon given by lattice QCD simulations \cite{LTY03,LTY04,TGLY04,AALTY10,HLY10}, namely the pion-nucleon sigma term and the chiral susceptibility, bringing severe constraints on their values~\cite{Chanfray2007,Ericson2007,Massot2008,Rahul,Cham1,Cham2}.
\end{enumerate}

In practice, the \textsl{chiral confining model} is implemented via a lagrangian whose explicit form is given for instance in Eq.~(1) of Ref.~\cite{Chanfray2024}.
It involves the scalar field $s$, i.e., the "nuclear physics sigma meson" $\sigma_W$ in our approach, the pion field $\phi_{a\pi}$, and the vector fields associated with the omega meson ($\omega^\mu$) and the rho meson ($\rho_a^\mu$) channels. Each meson-nucleon vertex is regularized by a monopole form factors (with respective cutoffs $\Lambda_{S,\pi,V,\rho}$) mainly originating from the nucleon compositeness, hence generating a bare OBE Bonn-like $NN$ interaction. For the generation of the G matrix this potential is completed by a fictitious $\sigma'$ meson exchange simulating the effect of two-pion exchange with $\Delta$'s in the intermediate state. This is a rather standard lagrangian but that takes advantage of being based on true coupling constants at finite density without ad hoc density dependent coupling constant. Instead, the three (multi)-body forces are generated by two specific crucial ingredients beyond the simplest approach and directly connected to the quark confinement mechanism and to the properties of the chirally broken vacuum: the introduction of an  in-medium modified nucleon mass which is supposed to embed its quark substructure and the presence of a chiral effective potential associated with the chirally broken vacuum.

\subsection{The in-medium nucleon mass}

The effective Dirac nucleon mass $M^*_N(s)$ deviates from the bare nucleon mass in presence of the nuclear scalar field $s$ as:
\begin{eqnarray}
M^*_N(s) &=& M_N + g_S\, s + \frac{1}{2}\kappa_{NS} \,s^2 + \mathcal{O}(s^3). 
\end{eqnarray}
In Refs.~\cite{Chanfray2005,Martini2006,Chanfray2007,Massot2008,Massot2009,Massot2012},  the pure linear sigma model has been considered   for the scalar coupling constant ($g_S=M_N/F_\pi$) but in the recent works \cite{Rahul,Cham1,Cham2,Universe,Chanfray2024}, $g_S$ was allowed to deviate from the L$\sigma$M and was fixed by a Bayesian analysis \cite{Rahul,Cham1,Cham2}. This quantity actually corresponds to the first order response of the nucleon to an external scalar field and can be obtained in an underlying microscopic model of the nucleon. The nucleon scalar  susceptibility $\kappa_\mathrm{NS}$ is another response parameter which reflects the polarization of the nucleon, i.e., the self-consistent readjustment of the quark wave function in presence of the scalar field.    Very generally the scalar coupling constant, $g_S$, and the nucleon response parameter,  $\kappa_{NS}$, depend on the quark substructure and the confinement mechanism as well as the effect of spontaneous chiral symmetry breaking. In our previous works \cite{Chanfray2005,Chanfray2007,Massot2008,Massot2009,Rahul,Massot2012} we introduced a dimensionless parameter, 
\begin{equation}
C\equiv \frac{\kappa_\mathrm{NS}\,F_\pi^2}{2 M_N},   
\end{equation}
which is expected to be of the order $C\sim 0.5$ as in the MIT bag  used in the QMC framework. Such a positive response parameter can be generated if confinement dominates spontaneous chiral symmetry breaking in the nucleon mass origin, as discussed in Ref.~\cite{Chanfray2011} within particular models.

The main purpose of this paper is to obtain an estimate of these response parameters, namely $g_S$ and $C$, from QCD phenomenology and more generally to obtain the evolution of the in-medium nucleon mass with increasing density.

\subsection{The scalar chiral effective potential}

In the majority of our previous works \cite{Martini2006,Chanfray2005,Chanfray2007,Massot2008,Massot2009,Rahul,Massot2012}, the chiral effective potential had the simplest linear sigma model (L$\sigma$M) form:
 \begin{equation}
V_{\chi,{L\sigma M}}(s)=\frac{1}{2}\,M^2_\sigma \,s^2\, +\,\frac{1}{2}\frac{M^2_\sigma -M^2_\pi}{ F_\pi}\, s^3\,+\,
\frac{1}{8}\,\frac{M^2_\sigma -M^2_\pi}{ F^2_\pi} \,s^4 \label{eq:VLSM}.
\end{equation} 
However for reasons given in Ref. \cite{Chanfray2023,Universe,Chanfray2024} and (re)explained below, we recently used an enriched chiral effective potential from a model able to give a correct description of the low-energy realization of chiral symmetry in the hadronic world. A good easily tractable candidate is the Nambu-Jona-Lasinio (NJL) model defined by the lagrangian given in Eq.~(9) of Ref.~\cite{Chanfray2023} or Eq.~(59) of Ref.~\cite{Universe}: it depends on four parameters: the coupling constants $G_1$ (scalar), $G_2$ (vector), the current quark mass $m$ and a (non-covariant) cutoff parameter $\Lambda$. Three of these parameters ($G_1$, $m$, and $\Lambda$) are adjusted to reproduce the pion mass, the pion decay constant and the quark condensate, ignoring pion-axial mixing.  We refer the reader to Refs.~\cite{Chanfray2011,Chanfray2023} for more details. As it was established in Ref.~\cite{Chanfray2023}, the net effect of the use of the NJL model is to replace the L$\sigma$M chiral potential by its NJL equivalent which is very well approximated by its expansion to third order in $s$:
\begin{equation}
V_{\chi,\mathrm{NJL}}(s)= \frac{1}{2}\,M^2_\sigma\, {s}^2\, +\,\frac{1}{2}\,\frac{M^2_\sigma -M^2_\pi}{ F_\pi}\, {s}^3\,\big(1\,-\,C_{\chi,\mathrm{NJL}}\big) +...\, .\label{eq:vchiNJL} 
\end{equation}
This version of the model described in  Refs.~\cite{Chanfray2023,Universe} has been called the \textsl{NJL chiral confining model}. For a given pion mass, sigma mass and pion decay constant parameters, the main difference with the original phenomenological version using the L$\sigma$M  lies in the presence of the $C_{\chi,\mathrm{NJL}}$ parameter whose expression in terms of a NJL loop integral is given in Ref.~\cite{Chanfray2023} and reminded in section 4 below. Considering typical values of the NJL parameters, we obtain values is in the range $C_{\chi,\mathrm{NJL}}: 0.4-0.5$.  The effect of the NJL model through the specific parameter $C_{\chi,\mathrm{NJL}}$ (named sometimes $C_\chi$ in the following) is thus to reduce the attractive tadpole diagram in the chiral potential, which makes it more repulsive.  
\subsection{Saturation mechanism and lattice QCD constraints}
\begin{figure}
\includegraphics[width=0.8\textwidth,angle=0]{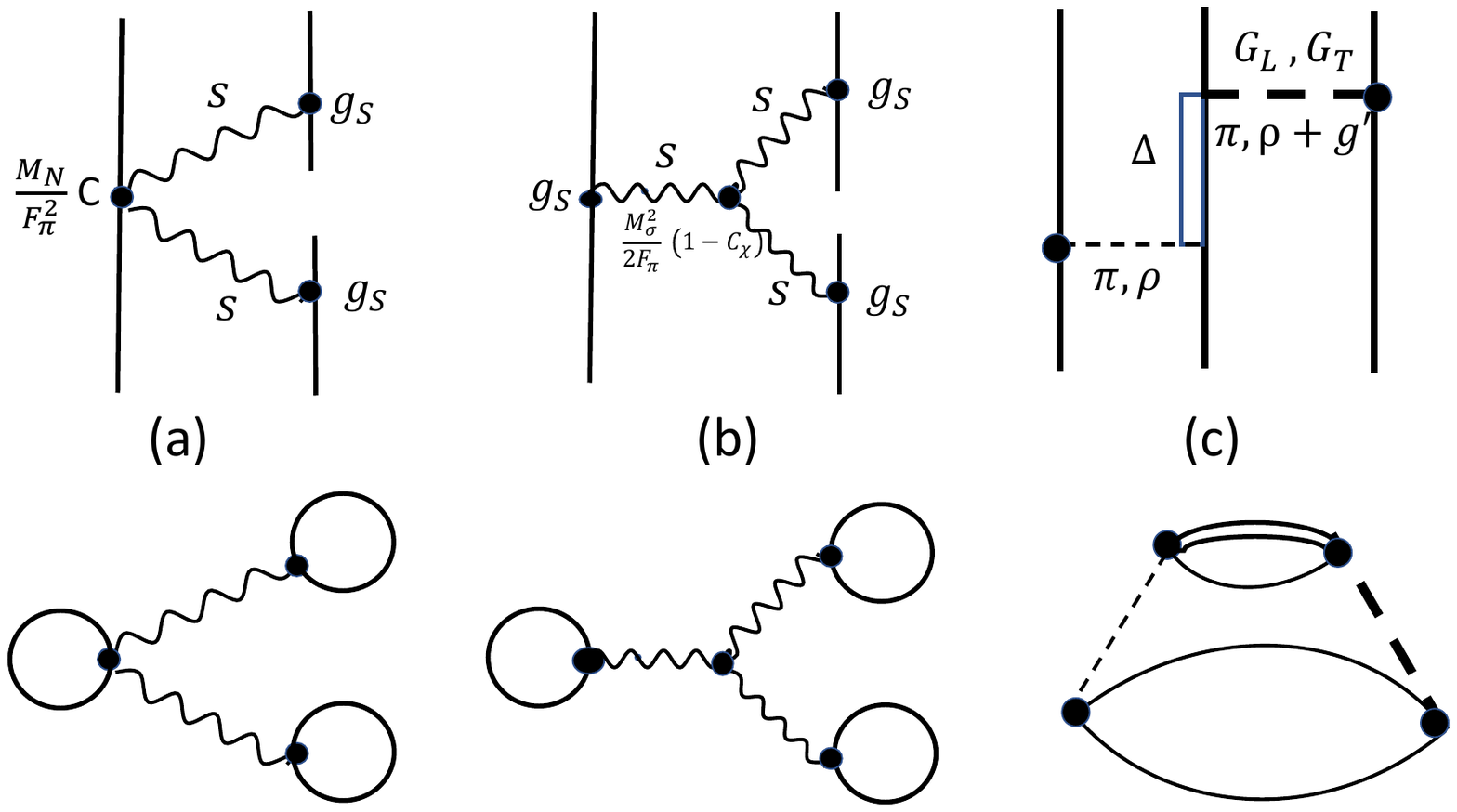}
\caption{Upper panel:  Nucleon three-body forces (3-hole-line) associated with the exchange of the scalar-isoscalar $s$ field in nuclear matter. (a) Response of the nucleon to the scalar field $s$ generating a repulsive three-body force; (b) tadpole diagram from the $s^3$ term in the chiral effective potential generating an attractive three-body force but corrected by the $C_\chi$ parameter; (c) another contribution to the three-body force presented in Ref.~\cite{Chanfray2024} coming from a Pauli-blocking effect in the $1-\Delta$ box corresponding to the earlier celebrated forces proposed in Refs. \cite{TB1,TB2,TB3,TB4,TB44} and also contained at NNLO in chiral EFT~\cite{CEFT}. Lower panel: the equivalent many-body diagrams. }
\label{THREEBODY}
\end{figure}

Combining the effects of the tadpole diagram  (i.e, associated with the $s^3$ in the chiral potential, see Fig.~\ref{THREEBODY}(b)) and of the response parameters (see Fig.~\ref{THREEBODY}(a)), it is possible to show that the scalar sector generates a three-nucleon contribution to the energy per nucleon, 
\begin{equation}
E^{(3b-s)}
=\frac{g^3_S}{2\,M^4_\sigma\,F_\pi}\,\left(2\, \frac{M_N}{g_S\,F_\pi}\,C -\,\left[1\,-\,C_\chi\right]\right)\,\rho^2_s\equiv\frac{g^3_S}{2\,M^4_\sigma\,F_\pi}\,\left(2\, \left[\frac{M_N}{g_S\,F_\pi}\,C +\,\frac{1}{2}\,C_\chi\right]\,-\,1\right)\,\rho^2_s,
\label{THREEBODN}
\end{equation}
where $\rho_s$ is the nucleonic scalar density, see Refs.~\cite{Ericson2007,Rahul,Chanfray2023,Chanfray2024,Universe,Cham2} for more details. If the confinement (represented by the parameter $C$ in Fig.~\ref{THREEBODY}(a))  dominates over the   chiral attraction (tadpole diagram shown in Fig.~\ref{THREEBODY}(b)), it  provides a very natural saturation mechanism but at variance with the QMC model which ignores the attractive tadpole diagram present in the chiral approach, this requires a value of the combination $C + C_\chi /2$  close to or even larger than one. In our first works, where the simplistic L$\sigma$M chiral potential ($C_\chi=0$) was used,  we were forced to conclude that a value of the $C$ parameter alone  close to or even larger than one is required~\cite{Chanfray2005,Chanfray2007,Massot2008,Massot2009,Rahul,Massot2012,Cham1}. In such a case the problem one has to face is that it seems impossible to find a realistic confining model for the nucleon able to generate $C$ larger than one~\cite{Chanfray2011,Chanfray2023}. In addition such a scenario generates a strong tension with lattice data discussed just below.

One very important point is that  these scalar response parameters can be related to some chiral properties of the nucleon. According to the lattice data  analysis of the Adelaide group, \cite{LTY03,LTY04,TGLY04,AALTY10}, the nucleon mass can be expanded in terms of the square of the pion mass, $M^2_{\pi}$, as $M_N(M^2_{\pi}) = 
a_{0}\,+\,a_{2}\,M^2_{\pi}\, +\,a_{4}\,M^4_{\pi}\,+ ...+\,\Sigma_{\pi}(M^2_{\pi},\, \Lambda)$ 
 where the pionic self-energy, $\Sigma_{\pi}(M^2_{\pi}, \Lambda)$, contains the non analytical contribution and is explicitly separated out. This latter term is calculated with just one adjustable cutoff parameter $\Lambda$ entering the $\pi NN, \pi N\Delta$ form factor regularizing the pion loops. While the $a_2$ parameter is related to the non pionic piece of the $ \pi N$ sigma term, the $a_4$ parameter is related to the nucleon QCD chiral susceptibility. The important point is that $a_4\simeq -0.5$~GeV$^{-3}$ is essentially compatible with zero in the sense that it is
much smaller than in a chiral effective model, $(a_4)_{L\sigma M} =-F_\pi\,g_{S}/2 M^4_{\sigma }\simeq -3.5$~GeV$^{-3}$, where the nucleon is seen as a juxtaposition of three constituent quarks getting their mass from the chiral condensate \cite{Ericson2007}. The explicit connection between the lattice QCD parameters $a_2$ and $a_4$, and the response parameter $g_S$ and $C$ in the L$\sigma$M case, has been first given in Refs.~\cite{Chanfray2007,Ericson2007,Massot2008}:  
\begin{equation}
a_2= \frac{F_\pi\, g_{S}}{M^2_{\sigma}}, 
\quad
a_4 = \frac{F_\pi\,g_{S}}{2 M^4_{\sigma }}\,\left(2\,\frac{M_N}{g_S\,F_\pi}\,C \,-\,3 \left[1\,-\,C_\chi\right]\right)\equiv\frac{F_\pi\,g_{S}}{2 M^4_{\sigma }}\,\left(2\,\left[\frac{M_N}{g_S\,F_\pi}\,C +\,\frac{3}{2}\,C_\chi\right]\,-\,3 \right).\label{LATTIX}
\end{equation}
Note that in the expression of $a_4$  the factor $M_N/F_\pi g_S$ in front of $C$, present in our recent papers \cite{Rahul,Cham1,Cham2,Universe,Chanfray2023,Chanfray2024}, was absent in \cite{Chanfray2007,Massot2008} since the nucleon mass was fixed to be $M_N=F_\pi g_S$. The parameter $C_\chi$ associated with the use of the NJL chiral effective potential has been introduced in \cite{Cham2,Chanfray2023,Chanfray2024,Universe}.  Depending of the details of the chiral extrapolation, the extracted values of the parameters are within the range in between  $a_2\simeq 1.5$~GeV$^{-1}$, $a_4\simeq -0.5$~GeV$^{-3}$~\cite{LTY04} and $a_2\simeq 1.0$~GeV$^{-1}$, $a_4\simeq -0.25$~GeV$^{-3}$~\cite{LTY03,AALTY10}. The quantity $a_2 M^2_\pi \sim 20-30$~MeV represents the non pionic piece of the sigma commutator directly associated with the scalar field $s$ (see the detailed discussion of this quantity in Ref.~\cite{Chanfray2007}).
One very robust conclusion is that the lattice result for $a_4$ is much smaller than the one obtained in  the simplest linear sigma model ignoring the nucleonic response ($C=0$) and the enriched form of the chiral potential ($C_\chi=0$), for which $a_4\simeq -3.5$~GeV$^{-3}$. Hence lattice data require a strong compensation from effects governing the three-body repulsive force needed for the saturation mechanism: compare Eqs.~(\ref{LATTIX}) and (\ref{THREEBODN}). In addition the compatibility of lattice data, nucleon modeling and saturation properties absolutely requires a non vanishing value of this $C_\chi$ parameter~\cite{Chanfray2023,Universe,Chanfray2024,Cham2}. One should also notice that the response parameters $g_S$ and $C$ entering Eq.~(\ref{LATTIX}) should not contain the pion cloud contribution when calculated in a given model, since in the lattice analysis the pionic self-energy, $\Sigma_{\pi}(M^2_{\pi}, \Lambda)$  is explicitly separated out. We will come to this point in the sixth section where the effect of the pion cloud on the response parameters will be discussed.

This model has been  applied in the past to the equation of state of nuclear matter~\cite{Chanfray2005,Chanfray2007,Massot2008,Massot2009,Rahul,Universe,Cham1,Cham2} and neutron stars~\cite{Massot2008,Massot2012,Cham1,Cham2} as well as to the study of chiral properties of nuclear matter~\cite{Chanfray2005,Martini2006,Ericson2007,Chanfray2007} at different levels of approximation in the treatment of the many-body problem (RMF, Relativistic Hartree-Fock or RHF, pion loop correlation energy). A short review is given in a recent paper~\cite{Chanfray2024} where in addition an improved calculation of the correlation energy is presented together with the idea of the "two sigma meson" picture ($s$ and $\sigma'$) each of them being  associated with  three-body forces; this later point is summarized on Fig. 1.  Finally let us stress again that the motivation of the present  paper  is   to provide a theoretical framework guiding our  phenomenological studies described above  where   direct comparison with empirical nuclear physics parameters are made. The first step  described in the next section concerns the  basic phenomenological QCD input.
\section{QCD phenomenology of the gluon correlator and effective quark interaction}

In this section, we  propose an effective model of low-energy QCD that incorporates both chiral symmetry breaking, i.e., the condensation of quark--antiquark pairs in the QCD vacuum,  and a confining string force that binds the massive constituent quarks inside the nucleon. It is based on the field correlator method (FCM) developed by Y. Simonov and collaborators \cite{Simonov1997,Simonov1998,Tjon2000,Simonov2002a,Simonov2002,Simonov-light,light1,light2,Digiacomo,Simonov2019}, where the starting point is the euclidean QCD lagrangian and partition function written with conventional notation:
\begin {eqnarray}
L_{\QCD} &=&  L_0 + L_1+ L_G, \qquad  Z = \int d\psi\, d\bar\psi\, dA_\mu \,e^{-\left[L_0 + L_1+ L_G\right]},\nonumber\\                                     \hbox{with}\quad L_0 &=&\int d^4x\,\bar\psi(x)\left(\gamma_\mu\partial_\mu +m\right)\psi(x),\nonumber\\
L_1 &=& \int d^4x\,\bar\psi(x)\, ig\gamma_\mu A_\mu \,\psi(x), \qquad
L_G  =  -\frac{1}{4}\int d^4x\, F^a_{\mu\nu}F^a_{\mu\nu}.
\end{eqnarray}
In the  first step we  integrate over the colored gluon fields ($A_\mu\equiv A^a_\mu\,t_a$) to generate a pure quark effective action. This is the gluon field averaging briefly described below. However, before carrying out this gluon integration, one makes a gauge choice (modified Fock-Schwinger gauge \cite{Simonov1997,Simonov1998,Tjon2000,Simonov2002a,Simonov2002,Simonov-light}) such that 
\begin{equation}
A_4\left(x_4,{\bf r}_0\right)=0,\,\,\,\,\left({\bf x}-{\bf r}_0\right)_j\cdot A_j\left(x_4,{\bf x}\right)=0,
\end{equation}
where ${\bf r}_0$ is an arbitrary point which will be identified later with a string junction coordinate. In this particular gauge, it is possible  to express $A_\mu$ in term of the field strength tensor $F_{\mu\nu}$ (\cite{Simonov1997,Simonov1998,Tjon2000,Simonov2002a,Simonov2002,Simonov-light}):
\begin{eqnarray}
A_4\left(x_4,{\bf x}\right)&=&\int_0^1 dv\,\left({\bf x}-{\bf r}_0\right)_k\, F_{k4}\left(x_4, {\bf r}_0+ v \left({\bf x}-{\bf r}_0\right)\right)\nonumber\\
A_j\left(x_4,{\bf x}\right)&=&\int_0^1 dv\,v\,\left({\bf x}-{\bf r}_0\right)_k\, F_{kj}\left(x_4, {\bf x}_0+ v \left({\bf x}-{\bf r}_0\right)\right).
\end{eqnarray}
Hence, the vector field appears as a contour integral, the contour $C({\bf x})$ reducing to a straight line ${\bf z}(v)={\bf r}_0+ v \left({\bf x}-{\bf r}_0\right)$ between the $x_4$ axis and the point where the gluon field is calculated. The partition function can be written in the successive forms,
\begin{eqnarray}
Z &=&\int d\psi\, d\bar\psi\, dA_\mu e^{-\left[L_0 + L_1+ L_G\right]}\equiv \int d\psi\, d\bar\psi\,e^{-L_0}\left\langle e^{-L_1}\right\rangle_G \nonumber\\
&\equiv & \int d\psi\, d\bar\psi\,e^{-\left[L_0 +L_1^{\eff}\right]}\equiv
\int d\psi\, d\bar\psi\,e^{-L^{\eff}},
\end{eqnarray}
where in the second form the gluon field averaging has been performed, which subsequently defines the effective quark lagrangian $L^{\eff}$, once gluon field has been integrated out. In the Field Correlator Method, the latter quantity can be obtained as an expansion of the exponential $\left\langle e^{-L_1}\right\rangle_G$, using the cluster expansion: \cite{Simonov1997,Simonov1998,Tjon2000,Simonov2002}, 
\begin{eqnarray}
 \left\langle e^{-L_1}\right\rangle_G &=&  \exp\left(\ln \left\langle e^{-L_1}\right\rangle_G \right)=\exp\left(\ln\left[1\, -\,\left\langle L_1\right\rangle_G + \frac{1}{2} \left\langle L_1 L_1\right\rangle_G \,-\, \frac{1}{6} \left\langle L_1 L_1 L_1\right\rangle_G + ... \right]\right)\nonumber\\
 &=&exp\bigg(-\left\langle L_1\right\rangle_G \,+\,\frac{1}{2}\,\big(\left\langle L_1 L_1\right\rangle_G\, -\,\left\langle L_1\right\rangle_G^2\big)\,  -\,\frac{1}{6}\,\big(\left\langle L_1 L_1 L_1\right\rangle_G\, -\, 3\,\left\langle L_1 L_1\right\rangle_G\left\langle L_1\right\rangle_G\nonumber\\ 
 &&+\,2\,\left\langle L_1\right\rangle_G^3\big) +...\bigg).
\end{eqnarray}
$L^{\eff}$ can be written as an infinite sum containing averages such as $\langle(A_\mu)^k\rangle$. According to Ref. \cite{Simonov2002}, one can make a Gaussian approximation, neglecting all correlators 
$\langle(A_\mu)^k\rangle$ of degree higher than $k=2$. The numerical accuracy has been studied on the lattice demonstrating that the corrections to the gaussian approximation are not larger than $3\%$ \cite{Simonov2002,Shev2000}.
Hence, neglecting all higher correlators this effective lagrangian (or more precisely this effective action) reads
\begin{eqnarray}
L^{\eff}&\simeq& L_0 +\frac{1}{2}\left\langle L_1 L_1 \right\rangle_G = \int  d^4x\,\bar\psi(x)\left(\gamma_\mu\partial_\mu \,+\, m\right)\psi(x)\nonumber\\
&& +\frac{1}{2}\int d^4x\,d^4y\,\,\,\bar\psi(x)\, \gamma_\mu\, t_a \,\psi(x)\,\,\,\,\bar\psi(y)\, \gamma_\nu\, t_b\, \psi(y)
\,\,\,\,\frac{\delta_{ab}}{C_F}\,J^{\mu\nu}(x,y),
\end{eqnarray}
with $C_F=(N^2_c -1)/2 N_c=4/3$ and
\begin{eqnarray}
&&J^{\mu\nu}(x,y)=\frac{g^2}{2N_c}\left\langle A_a^\mu(x) A_a^\nu(y)\right\rangle\nonumber\\
&=&\int_0^1 dv\,\int_0^1 dw\,\alpha_\mu(v)\,\alpha_\nu(w)\,\left({\bf x}-{\bf r}_0\right)_k\,\left({\bf y}-{\bf r}_0\right)_q\,\frac{g^2}{2N_c}\left\langle F_a^{k\mu}\left(z(v)\right)\,F_a^{q\nu}\left( z'(w)\right)\right\rangle ,
\end{eqnarray}
with $z_4=x_4$, ${\bf z}(v)={\bf r}_0+ v \left({\bf x}-{\bf r}_0\right)$,
$z'_4=y_4$, ${\bf z}'(w)={\bf r}_0+ w \left({\bf y}-{\bf r}_0\right)$, $\alpha_4(v)=1$, 
$\alpha_k(v)=v$.
This expression of the kernel can be shown to be gauge invariant, the reason being that the parallel transporters on the contour $C(x,y)$ are identically equal to unity in this gauge. The whole non perturbative physics is contained in the non local gluon condensate $g^2\left\langle F_a^{k\mu}\left(z(v)\right)\,F_a^{q\nu}\left( z'(w)\right)\right\rangle$ and  parametrizations are known from lattice measurements \cite{Digiacomo}. As in many previous works of Simonov and collaborators we keep only for the moment the non perturbative confining piece,
\begin{equation}
\frac{g^2}{2N_c}\left\langle F_a^{\rho\mu}(z)\,F_a^{\lambda\nu}(z')\right\rangle=\left(\delta_{\rho\lambda}\delta_{\mu\nu}-\delta_{\rho\nu}\delta_{\mu\lambda}\right) D(z-z'),
\end{equation}
where $D(x)$  describes the profile of the bilocal correlator of the non perturbative gluonic fields in the QCD vacuum. It follows immediately that its value at zero relative distance is directly related to the gluon condensate $\mathcal{G}_2$:
\begin{equation}
 D(0)=\frac{\pi^2}{18}\left\langle\frac{g^2}{4 \pi^2}\,F_a^{\mu\nu}\,F_a^{\mu\nu}\right\rangle  \equiv 
 \frac{\pi^2}{18}\,\mathcal{G}_2 .\label{GLUC}
\end{equation}
An additional information of this profile can be obtained from the properties of the averaged Wilson loop $W(C)$. Using the non abelian Stokes theorem and cluster expansion one has with self-evident notations
\begin{eqnarray}
 W(C)&=&\frac{1}{N_c}  \left\langle Tr\,P\,exp\left( ig\,\int_C dz_\mu\,A_\mu(z)\right)\right\rangle \nonumber\\
 &=& \frac{1}{N_c}  \left\langle Tr\,P\,exp\left(i g\,\int_S d\sigma_{\mu\nu}\,F_{\mu\nu}\right)\right\rangle \nonumber\\
 &=&\frac{1}{N_c}   Tr\,exp\left(-\frac{g^2}{2}\int_S d\sigma_{\mu\nu}\,d\sigma_{\lambda\rho}\,\left\langle F_{\mu\nu}F_{\lambda\rho}\right\rangle + ...\right) ,
\end{eqnarray}
where the last form is valid again under the assumption of the dominance of the bilocal correlator.
If we consider a particular Wilson loop with a rectangular contour $C_Q$ of spatial length R and euclidean time length T, one has the well known area law, 
\begin{equation}
 W(C)=\frac{1}{N_c}  \left\langle Tr\,P\,exp\left( ig\,\int_{C_Q} dz_\mu\,A_\mu(z)\right)\right\rangle = exp\left(-V_Q(R) T\right) =exp\left(-\sigma R T\right) =exp\left(-\sigma S_Q\right), 
\end{equation}
where $V_Q(R)=\sigma R$ is the interquark potential for heavy static quarks defining the string tension $\sigma$. Comparing the two expressions of the average Wilson loop one reaches the important result:
\begin{equation}
\sigma =\frac{1}{2} \int d^2 u\,D(u). \label{SR}
\end{equation}
We adopt a convenient gaussian parametrization \cite{Tjon2000,Simonov2002}: 
\begin{equation}
D(x)=D(0) \, e^{-x^2/4 T^2_g}\,\,\,\,\hbox{with}\,\,\,\,\,D(0)=\frac{\pi^2}{18}\,\mathcal{G}_2 \equiv\frac{\sigma}{2\pi T^2_g}\,\,\,\,\hbox{hence}\,\,\,\,\,T_g= \left(\frac{9}{\pi^3}\right)^{1/2}\left(\frac{\sigma}{\mathcal{G}_2}\right)^{1/2}.
\label{PROFIL}
\end{equation}
The quantity $T_g$ is the gluon correlation length which corresponds physically to the string radius. This form  of the bilocal correlator, which has the advantage of simplifying the calculations,  has been justified in \cite{Tjon2000,Simonov2002}. In short, since all the observables are integrals of $D(x)$, its explicit form is not essential at large
distances, provided it has a finite range $T_g$ and  the string tension, i.e., the coefficient in the area law of the Wilson loop, is equal to: $\sigma=\int d^2u\, D(u)/2$. Numerically the gluon correlation length can be estimated as:
\begin{equation}
T_g=0.336 \,\left(\frac{\sigma\,((\mathrm{GeV}^2)}{0.2}\right)^{1/2}\left(\frac{0.02}{\mathcal{G}_2\,(\mathrm{GeV}^4)}\right)^{1/2}\, \mathrm{fm}.
\end{equation}
A useful adimensional parameter, $\eta^2$, also used as a "small" parameter in the following, is:
\begin{equation}
\eta^2= \sigma\,T^2_g =
0.2\,\left(\frac{\sigma\,(\mathrm{GeV}^2)}{0.2}\right)\left(\frac{T_g\,(\mathrm{fm})}{0.1973}\right)^{2}\,=\,0.462\,\left(\frac{\sigma\,(\mathrm{GeV}^2)}{0.2}\right)\left(\frac{T_g\,(\mathrm{fm})}{0.3}\right)^{2}.
\end{equation}
As in Ref. \cite{Tjon2000,Simonov2002} we ignore the magnetic piece of the kernel, keeping only the dominant  electric piece $J_{44}$. In addition we make a static approximation, ignoring retardation effects:
\begin{equation}
e^{-x_4^2/4 T^2_g}\simeq 2\sqrt{\pi}\, T_g \,\delta (x_4).
\end{equation}
This approximation, already made in Ref. \cite{BBRV98} in the context of the heavy-light quark system, can be valid if the energy scale $T_g^{-1} \sim 700-800\,\mathrm{MeV}$ is bigger than the other scale $\sqrt{\sigma}\sim 400\,\mathrm{MeV}$ of the problem, or equivalently if the parameter $\eta$ is smaller than one.  Hence, we end up with an effective static lagrangian and consequently a more tractable effective hamiltonian governing the dynamics of light quarks moving in a BCS vacuum in presence of  a three-quarks junction  associated with the presence of a nucleon (or a static heavy quark) placed at ${\bf r}_0$ in the QCD vacuum. In principle the point ${\bf r}_0$ is arbitrary and should be chosen as the one minimizing the string lengths (Torricelli point) joining  the three quarks (see the discussion after Eq. (28) of \cite{Simonov2002}). In practice in the nucleon case we take it as a constant parameter coinciding with the three-quarks string junction and the three-quarks wave function is simply expressed in a factorized form in terms of single quark orbitals centered in ${\bf r}_0$ \cite{Simonov2002}. The underlying physical picture can be summarized in the statement that the three-quarks string, keeping the quarks together, is constructed on top of a chirally broken vacuum building a constituent quark at the end of the strings \cite{BBRV98}. Indeed the effective  hamiltonian reads : 
\begin{equation}
H=\int d^3x\,\psi^\dagger({\bf x})\left(-i\,\vec{\alpha}\cdot\vec{\nabla}\,+\, m\right)\psi({\bf x})
\,+\,\frac{1}{2}\int d^3x\,\int d^3y\,\psi^\dagger({\bf x})\,t_a\,\psi({\bf x})\,V\left({\bf x},{\bf y}\right)\,
\psi^\dagger({\bf y})\,t_a\,\psi({\bf y}).
\end{equation}
Introducing  ${\bf R}=({\bf x+y})/2-{\bf r}_0$, ${\bf r}={\bf x-y}$, ${\bf X}={\bf x}-{\bf r}_0$ and ${\bf Y}={\bf y}-{\bf r}_0$, the potential $V$, issued from $J_{44}$ \cite{Tjon2000,Simonov2002}, can be written in various forms:
\begin{eqnarray}
V\left({\bf x},{\bf y}\right)&=&\frac{1}{C_F}\,\frac{\sigma}{\sqrt{\pi}\,T_g}\,\left({\bf x}-{\bf r}_0\right)\cdot\left({\bf y}-{\bf r}_0\right)\,
\int_0^1 dv\,\int_0^1 dw\,e^{-\frac{\left(v\left({\bf x}-{\bf r}_0\right)-w\left({\bf y}-{\bf r}_0\right)\right)^2}{4 T^2_g}}\equiv \nonumber\\
V\left({\bf R},{\bf r}\right)&=&\frac{1}{C_F}\,\frac{\sigma}{\sqrt{\pi}\,T_g}\,\left({\bf R}^2-\frac{{\bf r}^2}{4}\right)\,\int_0^1 dv\,\int_0^1 dw\,e^{-\left(\frac{\left(v+w\right)\bf r}{4T_g}+
\frac{\left(v-w\right){\bf R}}{2T_g}\right)^2}\equiv\nonumber\\
V\left({\bf X},{\bf Y}\,;\,{\bf r}\right)&=&\frac{1}{C_F}\,\frac{\sigma}{\sqrt{\pi}\,T_g}\,\left(\frac{{\bf X}^2 +{\bf Y}^2}{2}-\frac{{\bf r}^2}{2}\right)\,I\left({\bf X},{\bf Y}\,;\,{\bf r}\right)\nonumber\\
\hbox{with}&&\quad I\left({\bf X},{\bf Y}\,;\,{\bf r}\right)\,=\,\int_0^1 dv\,\int_0^1 dw\,e^{-\left(\frac{\left(v+w\right)\bf r}{4T_g}+
\frac{\left(v-w\right)}{2T_g}\frac{{\bf X}+{\bf Y}}{2}\right)^2}.
\label{kernel}
\end{eqnarray}
The relative variable ${\bf r}$ is the one associated with the self-interaction of the quark and the variable ${\bf R}$ corresponds physically to the length of the confining string. We see (Eq. \eqref{kernel}) that in practice the self-interacting part and the string-like confining piece of the kernel might be mixed up in a very complicated way in the case of the presence of an "`external"' source, either a string junction or a static heavy (anti)quark. The treatment of the  chiral symmetry breaking versus confinement entanglement is certainly the most prominent problem of QCD which manifests already  at the level of the bare nucleon as well as at the level of dense and hot nuclear or hadronic matter. However some approximation schemes have been developed in the literature to partially disentangle these two aspects \cite{BBRV98,KNR2017}. Inspired by the third writing of the potential in Eq. (\ref{kernel}), we propose a specific ansatz interpolating between the ansatz used in Ref. \cite{BBRV98} and Ref. \cite{KNR2017} to realize it,
\begin{equation}
V\left({\bf X},{\bf Y}\,;\,{\bf r}\right)= V_C({\bf X})\,+\,V_C({\bf Y})\,+\,V^{NP}_{CSB}({\bf r}),  \label{STRUCT}    
\end{equation}
with
\begin{eqnarray}
V^{NP}_{CSB}({\bf r})&=&-\frac{1}{C_F}\,\frac{\sigma}{2\,\sqrt{\pi}\,T_g}\,{\bf r}^2
\,I\left(0\,,0\,;\,{\bf r}\right) \,+\,\frac{1}{C_F}\,\frac{4\,\sigma\,T_g}{\sqrt{\pi}}\\
\nonumber
&=&\frac{1}{C_F}\,\frac{2\,\sigma\,T_g}{\sqrt{\pi}}\,\left(2\,-\,\frac{{\bf r}^2}{4\,T^2_g}\,\int_0^1 dv\,\int_0^1 dw\,e^{-\left(\left(v+w\right)^2\frac{{\bf r}^2}{16 T^2_g}\right)}\right)\equiv\frac{1}{C_F}\,\frac{4\,\sigma\,T_g}{\sqrt{\pi}}\,v_l(r/T_g)
\end{eqnarray}
\begin{eqnarray}
V_C({\bf X})&=&\frac{1}{C_F}\,\frac{\sigma}{2\,\sqrt{\pi}\,T_g}\,{\bf X}^2
\,I\left({\bf X},{\bf X}\,;\,0\right)\,-\,\frac{1}{C_F}\,\frac{2\,\sigma\,T_g}{\sqrt{\pi}}\nonumber\\
&=&\frac{1}{C_F}\,\frac{\sigma}{2\,\sqrt{\pi}\,T_g}\,{\bf X}^2
\,\int_0^1 dv\,\int_0^1 dw\,e^{-\left(\left(v-w\right)^2\frac{{\bf X}^2}{4 T^2_g}\right)}\,-\,\frac{1}{C_F}\,\frac{2\,\sigma\,T_g}{\sqrt{\pi}},\label{VCONF}
\end{eqnarray}
where the constant term, $2U_0/C_F=4 \sigma T_g/C_F \sqrt{\pi}$, has been removed from the CSB piece and added to the confining piece, $V_C$, to have a vanishing $V^{NP}_{CSB}$ at infinity. The function $v_l(r)$ is a rapidly  decreasing function of $r$ with $v_l(0)=1$ and $v_l(\infty)=0$. Hence the effective interaction hamiltonian can be split according to: 
\begin{eqnarray}
H_{int}&=&H^{NP}_{CSB}\,+\,H_C\label{HINT}\\
H^{NP}_{CSB}&=&\frac{1}{2}\int d^3x\,\int d^3y\,\psi^\dagger({\bf x})\,t_a\,\psi({\bf x})\,V^{NP}_{CSB}\left({\bf r}\right)\,
\psi^\dagger({\bf y})\,t_a\,\psi({\bf y})\label{HCHI}\\
H_C&=&\frac{1}{2}\int d^3x\,\int d^3y\,\psi^\dagger({\bf x})\,t_a\,\psi({\bf x})\,K_C\left({\bf X}, {\bf Y}\right)\,
\psi^\dagger({\bf y})\,t_a\,\psi({\bf y})\nonumber\\
&&\hbox{with}\quad K_C\left({\bf X}, {\bf Y}\right)=V_C\left({\bf X}\right)\,+\,V_C\left({\bf Y}\right)\label{HCONF}.
\end{eqnarray}
The quantity $H^{NP}_{CSB}$ represents a non perturbative short-range  interaction generating chiral symmetry breaking and the condensation of light $q \bar{q}$ pair in the QCD vacuum,  whereas $H_C$ represents a long range interaction confining quarks inside the nucleon. In the following the QCD vacuum and the "shifted" vacuum at finite density, $\rho$, will be determined variationally as a BCS-like state $\left|\varphi(\rho)\right\rangle$. It is simple matter to check (see subsection 5.1) that $\langle\varphi(\rho)|\,H_C\,|\varphi(\rho)\rangle =0$. Hence the confining interaction in the presence of a string junction does not contribute to the chiral effective potential. Since this confining interaction exists only in a presence of an external source, i.e., a string junction, we  consider that it acts only on the valence quarks and not on the Dirac sea quarks.  The same mechanism  was described in Ref. \cite{BBRV98} for the heavy-light quark system with a heavy quark seen as an external source located in ${\bf r}_0$. Before going further let us remind some details given in \cite{Universe} about these two kernels.
\subsubsection{The self-energy kernel}
The short range ${\bf r}$-dependent piece $H_{CSB}$ will generate dynamical chiral symmetry breaking.  One finds analytically that $\int d^{3}r\, v_l(r/T_g)=\pi^{3/2}\, T^3_g\, \mu^{3}$ with $\mu^{3}=30$.
In \cite{Universe} we have shown (see Fig. 5 of this paper) that $v_l$ is well approximated by the Gaussian form:
\begin{equation}
 v^G_l(r/T_g)=e^{-\frac{r^2}{\mu^2\,T^2_g}}= e^{-\left(\frac{r}{3.1\,T_g}\right)^2}. 
\end{equation}
The momentum space representation of this interaction can be written as: 
\begin{eqnarray}
\hat{V}^{NP}_{CSB}\left({\bf q}\right)&=&\frac{1}{C_F}\,4\pi\,\sigma\,T^4_g\,\mu^3\,\Gamma ({\bf q})=\frac{1}{C_F}\,4\,N_c\,N_f\,G_1\Gamma ({\bf q})\label{NJLQ0}\\
\Gamma ({\bf q})&=&\frac{\int d^{3}r\,exp\left(-i{\bf q}\cdot{\bf r}\right)\, v_l(r/T_g)}{\int d^{3}r\, v_l(r/T_g)}
\simeq exp\left(-\frac{\mu^2\,T^2_g\,q^2}{4}\right)\label{NJLQ}.
\end{eqnarray}
At this point, a remark is in order. As pointed out in Ref. \cite{BBRV98}, this type of approach allows to establish a very appealing  connection between the gluon condensate and the light quark condensate, $\langle \bar{q} q\rangle = - C_{QG}\,T_g\,\mathcal{G}_2$ (see Eq. (22) of  \cite{BBRV98}). We obviously also have a direct connection  between the equivalent NJL $G_1$ parameter discussed in the next section and the gluon condensate (see Eqs.  \ref{GLUC},\ref{PROFIL},\ref{NJLQ0},\ref{G1NAIVE}). Accordingly we also have a direct connection between the gluon condensate and the constituent quark mass and  therefore the light quark condensate, although this relationship is different in  details from Eq. (22) of  \cite{BBRV98}. Here we simply note that with the set of parameters used in our previous paper \cite{Universe} (and corresponding to the NJLset1 set of parameters introduced below), the numerical value of the $C_{QG}$ constant  is  $C_{QG}=0.39$ to be compared with Eq. (145) of Ref. \cite{Simonov1997} and Eq. (36) of Ref. \cite{Simonov1998} which suggests a value of $C_{QG}$  of order unity.  Also note that, in practice, replacing the Gaussian fall-off by a more realistic exponential fall-off of the profile $D(x)$ will certainly (slightly) modify the equivalent NJL  $G_1$ parameter and consequently the relationship between the gluon condensate and the light quark condensate. It remains nevertheless true that the light quark condensate is related in our approach only to the chromoelectric part of the gluon condensate in agreement with the fact that  the chiral condensate vanishes beyond the critical  temperature while the chromo-magnetic part of the gluon condensate does not. This latter point is however beyond the scope of this paper.
\subsubsection{The confining kernel}
The confining interaction is given by Eq. (\ref{VCONF}). We see immediately that it has a quadratic behavior at  short distance, ($R<<T_g$), whereas at large distance it is possible to show that it has a linear behavior: 
\begin{eqnarray}
R<<T_g & :&  \qquad (V_C)_S\left({\bf\ R}\right)+\frac{1}{C_F}\,\frac{2\,\sigma\,T_g}{\sqrt{\pi}}=\frac{1}{C_F}\frac{\sigma}{2\,\sqrt{\pi}\,T_g}\, R^2\nonumber\\
R>>T_g & :&  \qquad (V_C)_L\left({\bf R}\right)+\frac{1}{C_F}\,\frac{2\,\sigma\,T_g}{\sqrt{\pi}}=\frac{1}{C_F} \,\sigma\,\left(R\,-\,\frac{2}{\sqrt{\pi}}\,T_g\right).
\end{eqnarray}
For bound state calculation, in order  to simplify the numerical computation we make use of an approximate expression interpolating between the short range and the long range behaviours: 
\begin{eqnarray}
 &&(V_C)_I\left({\bf R} \right) =F(R)\,(V_C)_S\left({\bf R}\right)\,+\,\left(1-F(R)\right)(V_C)_L\left({\bf R}\right).\nonumber\\
 &&\hbox{with}\qquad  F(R)=\left[1\,+\,exp\left(\frac{R-(6/\sqrt{\pi})\,T_g}{0.1\,T_g}\right)\right]^{-1}.
\end{eqnarray}
This confining interaction (without the constant shift) is displayed in Figures 6 and 7  of \cite{Universe} for typical values of the
parameters. This constant shift  generates an attractive potential well of depth $2 \sigma T_g/C_F \sqrt{\pi}\sim 250\,\mathrm{MeV}$. Such an attractive  pocket is particularly welcome to generate a not too large nucleon mass and is frequently included in alternative confining potential \cite{Jena97}.
\section{Breaking of chiral symmetry and the gap equation}
The general picture underlying our approach can be summarized as follows.   Nuclear matter is made of nucleons  which look like  Y-shaped strings  generated by a non perturbative confining force, with constituent quarks at the ends moving in the chirally broken vacuum.
The building of the nucleon state and the confined quarks wave functions  using the above confining interaction, $V_C(R)$, is postponed to section 5. The object of this section is to study the ability of our approach to generate a chirally broken vacuum reproducing some basic vacuum properties such as the constituent quark mass, the quark condensate, the pion decay constant and the pion mass.\\

We first give a general discussion of this CSB mechanism using the above non perturbative potential $\hat{V}^{NP}_{CSB}\left({\bf q}\right)$ possibly completed by perturbative  gluon exchange contribution $\hat{V}^{PG}_{CSB}\left({\bf q}\right)$. For that purpose we consider a very general (instantaneous) color interaction, 
\begin{equation}
\hat{V}_{CSB}\left({\bf q}\right)=   \left(\hat{V}_0\left({\bf q}\right)\,\gamma_0\,\gamma_0\,-\,\hat{V}_g\left({\bf q}\right)\,\bfvec{\gamma}\cdot\bfvec{\gamma}\,+\,\hat{V}_q\left({\bf q}\right)\,\bfvec{\gamma}\cdot\hat{\bf q}\,\bfvec{\gamma}\cdot\hat{\bf q}\right)\,t_a \,t_a.
\end{equation}
Very generally, the quark field operator  for a given flavor and  color can be expanded on a BCS basis, according to:
\begin{equation}
q({\bf{r}},t)=\frac{1}{\sqrt{V}}\,	\sum_k\,\eta_k\,C_k\,e^{i\,\bf{k}\cdot\bf{r}}.\label{QEXP}
\end{equation}
We adopt for convenience a discrete normalization and $V$ is the volume of the large box. The $C_k$'s are positive and negative energy state annihilation operators representing either a true destruction operator of a constituent quark or a true creation operator of an antiquark.
The states labeled by  $k=({\bf{k}}, s, \varepsilon)$  correspond to a plane wave basis: $\bf{k}$ is the  momentum, $s=\pm 1$ represents the polarization state and $\varepsilon$ allows to distinguish between positive and negative energy states:
\begin{eqnarray}
& &\varepsilon=+1,\quad C_{\bf{k}, s, +1}=B_{\bf{k}, s}\, ,\quad \eta_{\bf{k}, s, +1}=u(\bf{k},  s)\nonumber\\
& &\varepsilon=-1,\quad C_{\bf{k}, s, -1}=D^\dagger_{-\bf{k}, -s}\, ,\quad \eta_{\bf{k}, s, -1}=v(-\bf{k}, -s).
\end{eqnarray} 
The explicit forms of the the Dirac spinors in term of the  chiral angles ($s_k=\sin\varphi_k, c_k=\cos\varphi_k$) are: 
\begin{equation}
\eta_{\bf{k}, s, +1}=u({\bf k}, s)=\sqrt{\frac{1}{2}}\left(
\begin{array}{c}
                \sqrt{1+s_k}\,\,\chi_s \\
                \sqrt{1-s_k}\,\,\bfvec{\sigma}\cdot {\bf k}\,\,\chi_s
\end{array}
\right)
\end{equation}
\begin{equation}
\eta_{\bf{k}, s, -1}=v(-{\bf k}, -s)=\sqrt{\frac{1}{2}}\left(
\begin{array}{c}
                -\sqrt{1-s_k}\,\,\bfvec{\sigma}\cdot {\bf k}\,\,\chi_s \\
                \sqrt{1+s_k}\,\,\chi_s 
\end{array}
\right).
\end{equation}
The positive energy and negative energy projectors are: 
\begin{equation}
\Lambda_{+}({\bf k})=\sum_s  \eta_{\bf{k}, s, +1}\,\bar\eta_{\bf{k}, s, +1} =\frac{\gamma_0\, - \,c_k\,\bfvec{\gamma}\cdot{\bf{k}}\, +\,s_k }{2} \equiv \frac{\gamma_0 }{2}-\Lambda_{red}({\bf k})
\end{equation}
\begin{equation}
\Lambda_{-}({\bf k})=\sum_s  \eta_{\bf{k}, s, -1}\,\bar\eta_{\bf{k}, s, -1} =\frac{\gamma_0\, + \,c_k\,\bfvec{\gamma}\cdot{\bf{k}}\, -\,s_k }{2} \equiv \frac{\gamma_0 }{2}+\Lambda_{red}({\bf k}). \label{PROJ}
\end{equation}
The inverse quark propagator has the form:
\begin{equation}
 S^{-1}({\bf p}) =\gamma_0 \,p_0\,- \,\bfvec{\gamma}\cdot{\hat{\bf p}}\,B(p)\,+\,A(p).
\end{equation}
The connection with the chiral phase is:
\begin{equation}
s_p=\sin\varphi_p=\frac{A(p)}{\sqrt{A^2(p)\,+\,B^2(p)}}\equiv\frac{A(p)}{\varepsilon_p},\quad
c_p=\cos\varphi_p=\frac{B(p)}{\sqrt{A^2(p)\,+\,B^2(p)}}\equiv\frac{B(p)}{\varepsilon_p}.
\end{equation}
Minimization of the vacuum Hartree-Fock energy yieds the (renormalized) Dyson-Schwinger equation
\begin{eqnarray}
 A(p)&=&Z_m\,m_0\,+\,\frac{C_F}{2}\,\int \frac{d^3k}{(2\pi)^3}\,\bigg[\left(\hat{V}_0\,+\,3\,\hat{V}_g\,-\,\hat{V}_q\right)({\bf p}-{\bf k})\,\frac{A(k)}{\varepsilon_k}\bigg] \\
 B(p)&=&Z_2\,p\,+\,\frac{C_F}{2}\,\int \frac{d^3k}{(2\pi)^3}\,\bigg[\left(\hat{V}_0\,+\,\hat{V}_g\,-\,\hat{V}_q\right)({\bf p}-{\bf k})\,\hat{\bf p}\cdot\hat{\bf k} \nonumber\\
& & \qquad\qquad\qquad\qquad +\, 2\,\hat{V}_q({\bf p}-{\bf k})\,\hat{\bf q}\cdot\hat{\bf p}\,\hat{\bf q}\cdot\hat{\bf k}\bigg]\,\frac{B(k)}{\varepsilon_k}\bigg]_{{\bf q}={\bf p}-{\bf k}}, 
\end{eqnarray}
where $m_0$ is the bare current light quark mass and $Z_m$ and $Z_2$ are the mass and wave functions renormalization parameters. Let us introduce the following notations,
\begin{equation}
Z(p)=\frac{B(p)}{p},\quad M(p)=\frac{A(p)}{Z(p)}, \quad E_p=\sqrt{M^2(p)+p^2}=\frac{\varepsilon_p}{Z(p)},   
\end{equation}
such that:
\begin{equation}
s_p=\frac{A(p)}{\varepsilon_p}=\frac{M(p)}{E_p},\qquad    c_p=\frac{B(p)}{\varepsilon_p}=\frac{p}{E_p}.
\end{equation}
From the above Dyson-Schwinger equation, one obtains a gap equation whose solution is the momentum dependent constituent quark mass, $M(p)$:
\begin{equation}
M(p)=\frac{A(p)}{Z(p)}=\frac{Z_m\,m_0\,+\,\frac{C_F}{2}\,\Sigma_S(p)}{Z_2\,+\,\frac{C_F}{2}\,\Gamma_S(p)},    
\end{equation}
with
\begin{eqnarray}
 \Sigma_S(p)&=&  \frac{d^3k}{(2\pi)^3}\,\bigg[\left(\hat{V}_0\,+\,3\,\hat{V}_g\,-\,\hat{V}_q\right)({\bf p}-{\bf k})\,\frac{M(k)}{E_k}\bigg]\nonumber\\ 
 \Gamma_S(p)&=&\int \frac{d^3k}{(2\pi)^3}\,\bigg[\left(\hat{V}_0\,+\,\hat{V}_g\,-\,\hat{V}_q\right)({\bf p}-{\bf k})\,\hat{\bf p}\cdot\hat{\bf k}\nonumber\\
&& \qquad\qquad\qquad\qquad +\, 2\,\hat{V}_q({\bf p}-{\bf k})\,\hat{\bf q}\cdot\hat{\bf p}\,\hat{\bf q}\cdot\hat{\bf k}\bigg]\frac{(k/p)}{E_k}\bigg].
\end{eqnarray}
The renormalization conditions are formulated at an arbitrary scale $\mu_R$:
\begin{eqnarray}
  M(p=\mu_R)&=&m(\mu_R)=Z_m\,m_0\,+\,\frac{C_F}{2}\,\Sigma_S(p=\mu_R)\\ Z(p=\mu_R)&=&1=\quad {Z_2\,+\,\frac{C_F}{2}\,\Gamma_S(p=\mu_R)}.    
\end{eqnarray}
Hence the renormalized gap equation finally reads:
\begin{equation}
M(p)=\frac{m(\mu_R)\,+\,\frac{C_F}{2}\,\left[\Sigma_S(p)\,-\,\Sigma_S(\mu_R)\right]}{1\,+\,\frac{C_F}{2}\,\left[\Gamma_S(p)\,-\,\Gamma_S(\mu_R)\right]}.     
\end{equation}
We will take in practice a renormalization point around $\mu_R\sim 2 GeV$, for which the running light quark mass is expected to be $m(\mu_R)\sim 4\,\mathrm{MeV}$. However in the following we will take the freedom to adjust this running mass (further identified with the quark mass parameter $m$ entering the NJL model) to get the correct pion mass. \\

We first consider the case where only the non perturbative piece (Eq. \ref{NJLQ}) is considered, namely:
\begin{equation}
\hat{V}_0(q)= \hat{V}^{NP}_{CSB}\left({\bf q}\right)=\frac{1}{C_F}\,4\pi\,\sigma\,T^4_g\,\mu^3\,\Gamma ({\bf q}),\qquad   \hat{V}_g(q)=\hat{V}_q(q)=0.
\end{equation}
A preliminary calculation (not solving fully the equation but using a simple functional ansatz for $M(p)$), taking typical values of the phenomenological QCD parameters, $\sigma = 0.18 \div 0.20\,GeV^2$, $\mathcal{G}_2 =0.02 \div 0.05\,GeV^4$ (or $T_g=0.2 \div 0.3 fm$), leads to  a non trivial solution of the gap equation but with a constituent quark mass, $M_0=M(p=0)$, not larger than $100\,\mathrm{MeV}$.  A constituent quark mass of the order of $300\, \mathrm{MeV}$ necessitates an increase of the cutoff $\Lambda_{NP}=2/\mu T_g$ by at least factor two. This implicit constraint on the constituent quark mass, namely $M_0>300\,MeV$, is motivated by the following reason:  we choose to fix the string tension to the value $\sigma=0.18\,GeV^2$ and to fix the remaining parameters (essentially $T_g$ or  alternatively $\mathcal{G}_2$) entering the CSB potential,  not only to generate a correct quark condensate, $\langle \bar{q} q\rangle \sim -(250 \,MeV)^3$,  and pion decay constant, $F_\pi\sim 90\,MeV$, but also to generate a constituent quark mass $M_0$ larger than  $300\,MeV$ in order to make contact with standard NJL phenomenology, the choice of the NJL approximation itself being motivated by the need to make the nuclear physics applications tractable, as discussed below.\\

From the previous discussions, it appears that the non perturbative (NP) CSB interaction is not able to generate  sufficient chiral symmetry breaking. This conclusion seems to contradict Refs. \cite{Simonov1997,Simonov1998,light1}  where the pure NP-CSB interaction is shown to be able to generate acceptable quark condensate and pion decay constant, using a value of the string radius not larger than $T_g=0.2 \,fm$. Actually using such a low value of $T_g$  in the equivalent NJL model described below, it is possible to generate phenomenologically acceptable values for the quark condensate and the pion decay, but at the cost of generating  a value of the constituent quark mass, $M_0\sim 200\,MeV$, significantly lower than $300\,MeV$. Hence, there is no real contradiction between our modeling and Refs. \cite{Simonov1997,Simonov1998,light1}.  It thus follows that if we are to maintain this constraint on the value of the constituent quark mass  some other contribution is required. A plausible candidate to be included on top of the non perturbative FCM interaction  is perturbative one gluon exchange to complement the pure NP-CSB one. To have a first idea of its effect, ignoring the  problem gauge of consistency, we consider the Landau gauge transverse gluon exchange from Ref. \cite{Maris} but taken in the static limit:
\begin{equation}
 \hat{V}^{PG}(q)=  \frac{1}{q^2} \,\frac{8\pi^2\,\gamma_m}{\ln \left(e^2\,-\,1\,+\,\left(1\,+\,q^2/\Lambda^2_{QCD}\right)^2\right)}\,\left(1\,-\,e^{-q^2/4 m^2_t}\right),
\end{equation}
with:
\begin{equation}
m_t=0.5\,GeV,\qquad  \gamma_m=\frac{12}{33\,-\,2\,N_f}=\frac{12}{29},\qquad  \Lambda_{QCD}=234\,\mathrm{MeV}.
\end{equation}
Hence the various Dirac component of $\hat{V}_{CSB}\left({\bf q}\right)$ are: 
\begin{equation}
\hat{V}_0(q)= \hat{V}^{NP}_{CSB}(q)\,+\,\hat{V}^{PG}(q),\qquad   \hat{V}_g(q)=\hat{V}_q(q)=\hat{V}^{PG}(q).
\end{equation}
In such a case and according to a first estimate,  it seems possible to generate a solution with $M_0$ close to $300\,\mathrm{MeV}$. Even if we admit that the consistent incorporation of the one-gluon exchange deserves a much more thorough study one can arrive at a qualitative but rather robust conclusion. Neither the non perturbative interaction nor the perturbative gluon exchange taken alone are able to generate sufficient chiral symmetry breaking. However adding the two contributions together seems to be sufficient. In addition the non perturbative contribution and the perturbative one contribute about equally to the generation of the constituent quark mass since $\hat{V}^{NP}_{CSB}(q=0)\sim 3\,\hat{V}^{PG}(q=0)$. Before going further,  we should stress that the above conclusion has to be confirmed by  modern studies based on the Dyson–Schwinger equations,  which suggest that the simple non-perturbative
one-gluon exchange should be replaced by a four-quark Green’s function, where the required $1/q^4$-behavior is achieved by a combination of the quark-gluon vertices with a non-singular gluon propagator \cite{Green}. \\

Our strategy to build the QCD-connected version of the chiral confining model is to perform a semi-bosonization of the lagrangian $\mathcal{L}_{CSB}$, i.e., associated to the CSB interaction, to generate the meson sector lagrangian and the coupling of the various meson fields to the valence quark sector. As explained in details in section 5, the presence of the confining kernel induced by the presence of the string junction, reduces
to modify the  action $S_{CSB}$ by just adding a static mass operator to the constituent quark propagator which allows to calculate the bound state wave function of the confined constituent quarks, the nucleon mass, all the response parameters such as $g_S$ or $\kappa_{NS}$ and  the Yukawa meson-nucleon coupling-constants. However, keeping the finite-range CSB interaction would imply the use of bilocal meson  fields  resulting from the bosonization procedure which was used in our previous  paper \cite{Chanfray2011}.  This cumbersome formalism  would have required completely reformulating and extending  our approach and would also lack its simplicity. Hence for practical reasons in view of nuclear physics applications we will simulate the BCS non local interaction by a point-like one of the NJL type characterized by four parameters: the coupling constants $G_1$ (scalar) and $G_2$ (vector), the current quark mass $m$ (identified with running quark mass at scale $\mu_R\sim 2\,GeV$) and a (non covariant) cutoff parameter $\Lambda$. 
\subsection{Set of NJL parameters NJLSET1}
In  the pure non perturbative case, we have shown in our previous paper \cite{Universe} that the non local kernel can be approximated  by a point-like interaction with the same total strength. By performing a Fierz transform \cite{Simonov2002a,Simonov-light,light1,light2} of the lagrangian or hamiltonian of the  type, $(\psi^\dagger t_a\psi )(\psi^\dagger t_a\psi )$, one can check that this interaction is equivalent to a Nambu--Jona--Lasinio (NJL) model  with a scalar sector coupling constant $G_1$ (we take $N_f=2,\, N_c=3)$,
\begin{equation}
G_1=\frac{4\pi\,\sigma\,T^4_g\,\mu^3}{4\,N_c\,N_f} = 3.1416\,  \left(\frac{\sigma\,(\mathrm{GeV}^2)}{0.2}\right)\left(\frac{T_g\,(\mathrm{fm})}{0.1973}\right)^{4}\, \mathrm{GeV}^{-2}, \label{G1NAIVE}
\end{equation}
with:
\begin{equation}
 \Lambda = \left(\frac{6\sqrt{\pi}}{\mu^3}\right)^{1/3} \,\frac{1}{T_g}=0.71\,\left(\frac{0.1973}{T_g(\mathrm{fm})}\right)\, \mathrm{GeV}.    
\end{equation}
Starting with accepted values of the string tension and gluon condensate,  we took in \cite{Universe,Chanfray2024} $\sigma=0.18\,\mathrm{GeV}^2$ and  $\mathcal{G}_2=0.025\,\mathrm{GeV}^4$, which implies that $T_g= 0.286\,\mathrm{fm}$, and hence, $\eta=\sqrt{\sigma}T_g=0.615$. Consequently,  the NJL parameters are $G_1=12.514 \,\mathrm{GeV}^{-2}$ and $\Lambda =0.488\, \text{GeV}$. It follows that $G_1\Lambda^2=3$ is just below the critical value for chiral symmetry breaking. Hence, the model limited to the BCS-NP kernel  is not able to properly generate the physical broken vacuum. We thus  decided to  increase the cutoff parameter to $\Lambda =0.604\,\mathrm{MeV}$ to obtain the correct phenomenology. We also found that the NJL parameter entering the cubic term of the chiral effective potential is significant, $C_\chi =0.488$, and the effective sigma mass parameter is $M_\sigma=716.4$~MeV. A calculation of the nuclear EOS with this set of parameters is presented in the last section of \cite{Chanfray2024}. In the following we will refer to this set of parameters as NJLSET1.
\subsection{Set of NJL parameters NJLSET2}
In the case where the perturbative gluon  exchange is included we start with a larger value of the gluon condensate, $\mathcal{G}_2=0.039 \,\mathrm{GeV}^4$. Keeping the same value of the string tension, $\sigma=0.18\,\mathrm{GeV}^2$, one obtains a smaller value of the string radius, $T_g= 0.229\,\mathrm{fm}$ and hence, $\eta=\sqrt{\sigma}T_g=0.492$. We also use a smaller value of the parameter $m_t=0.86\,\mathrm{GeV}$. By Fierz transform an equivalent NJL model can be generated with coupling constants which can be estimated as:
\begin{eqnarray}
 G_1 &=& \frac{C_F}{4\,N_c\, N_f}  \left(\hat{V}^{NP}_{CSB} (q=0)\,+\,4\,\hat{V}^{PG}(q=0)\right)=10\,\mathrm{GeV}^2\\ \label{G1NPG}
 G_2 &=& \frac{C_F}{4\,N_c\, N_f}  \left(\hat{V}^{NP}_{CSB} (q=0)\,+\,2\,\hat{V}^{PG}(q=0)\right)=7.5\,\mathrm{GeV}^2. \label{G2SPACE}
\end{eqnarray}
Using for the NJL light quark mass parameter, $m=4\,\mathrm{MeV}$ and a NJL cutoff, $\Lambda=0.665\
,\mathrm{GeV}$, we obtain (see next section) $M_0=358\,\mathrm{MeV}$, $F_\pi=86.7\,\mathrm{MeV}$ and 
$M_\pi=138.2\,\mathrm{MeV}$. We also found that the NJL parameter entering the cubic term of the chiral effective potential is significant, $C_\chi =0.458$. In the following we will refer to this set of parameters as NJLSET2.
\subsection{Lack of vector repulsion}
In reality, the $G_2$ parameter of Eq. (\ref{G2SPACE}) only concerns the space component of the vector interaction entering the NJL lagrangian of Eq. (\ref{LNJL}), and governing the pion-axial-vector mixing. This can be easily understood when considering the Fierz transforms of the colored interactions projected onto the the various components of the NJL interaction: 
\begin{eqnarray}
 &&\gamma_0\gamma_0\,t_a t_a\quad\to\quad \frac{C_F}{N_c N_f}  \bigg(-\frac{1}{4}\,\big(1 1 +i\gamma_5 i\gamma_5\big)\nonumber\\
 &&\qquad\qquad\qquad\qquad\qquad+\,\frac{1}{4}\big(-\gamma_0\gamma_0\,-\,\bfvec{\gamma}\cdot\bfvec{\gamma}\,-\,\gamma_0\gamma_5\,\gamma_0\gamma_5\,-\,
 \vec{\gamma}\gamma_5\cdot\vec{\gamma}\gamma_5\big)\bigg)\,\big(1+\vec\tau\cdot\vec\tau\big)\nonumber\\
&&-\bfvec{\gamma}\cdot\bfvec{\gamma} \,t_a t_a\quad\to\quad \frac{C_F}{N_c N_f}  \bigg(-\frac{3}{4}\,\big(1 1 +i\gamma_5 i\gamma_5\big)\nonumber\\
 &&\qquad\qquad\qquad\qquad\qquad+\,\frac{1}{4}\big(3\,\gamma_0\gamma_0\,-\,\bfvec{\gamma}\cdot\bfvec{\gamma}\,+\,3\,\gamma_0\gamma_5\,\gamma_0\gamma_5\,-\,
 \vec{\gamma}\gamma_5\cdot\vec{\gamma}\gamma_5\big)\bigg)\big(1+\vec\tau\cdot\vec\tau\big)\nonumber\\
 &&\gamma^\mu\cdot\gamma_\mu \,t_a t_a\quad\to\quad \frac{C_F}{N_c N_f}  \bigg(-\,\big(1 1 +i\gamma_5 i\gamma_5\big)\nonumber\\
 &&\qquad\qquad\qquad\qquad\qquad+\,\frac{1}{2}\big(\gamma_0\gamma_0\,-\,\bfvec{\gamma}\cdot\bfvec{\gamma}\,+\,\gamma_0\gamma_5\,\gamma_0\gamma_5\,-\,
 \vec{\gamma}\gamma_5\cdot\vec{\gamma}\gamma_5\big)\bigg)\big(1+\vec\tau\cdot\vec\tau\big).
\end{eqnarray}
One sees that the Coulomb-like interaction $\gamma_0\gamma_0\,t_a t_a$ generates a time component of the NJL vector interaction with the "wrong" sign, namely attractive and not repulsive. Consequently, the strength of the time component of the vector interaction is 
\begin{equation}
G_2^{(time)} = \frac{C_F}{4\,N_c\, N_f}  \left(-\hat{V}^{NP}_{CSB} (q=0)\,+\,2\,\hat{V}^{PG}(q=0)\right), \label{G2TIME}    
\end{equation}
which is very small or even negative, whereas one should generate a repulsive omega (and rho) exchange such that $G_2^{time}=g^2_V/M^2_V$ with $g_v\sim 2.65$. This is a real weak point of the approach and we have presently  no solution for this problem. In practice, as in our papers using a Bayesian analysis \cite{Rahul,Cham1,Cham2}, we consider the $\omega NN$ coupling constant as an independent free parameter. 
\section{The NJL model approximation for practical nuclear physics application}
\subsection{The bosonized NJL lagrangian}
The lagrangian of the NJL model in its original formulation is: 
\begin{eqnarray}
{\cal L}&=& \bar{\psi}\left(i\,\gamma^{\mu}\partial_\mu\,-\,m\right)\,\psi\,+\,\frac{G_1}{2}\,\left[\left(\bar{\psi}\psi\right)^2\,+\
\left(\bar{\psi}\,i\gamma_5\vec\tau\,\psi\right)^2\right]\nonumber\\
& &-\,\frac{G_2}{2}\,\left[\left(\bar{\psi}\,\gamma^\mu\vec\tau\,\psi\right)^2\,+\,
\left(\bar{\psi}\,\gamma^\mu\gamma_5\vec\tau\,\psi\right)^2\,+\,\left(\bar{\psi}\,\gamma^\mu\,\psi\right)^2\right]. \label{LNJL}
\end{eqnarray}
Using path integral techniques and after a chiral rotation of the quark field, it can be equivalently  written in a semi-bosonized form involving a pion field embedded in the unitary operator $U=\xi^2$, a scalar field, ${\cal S}$, a vector field, ${V}^\mu$, and an axial-vector field,  ${A}^\mu$. It has the explicit form (see  Eqs (2, 7-11) in Ref. \cite{Chanfray2011} and subsection 4.2 below for the details of the notations):
\begin{eqnarray}
{\cal L}&=& \bar{q}\left[i\,\gamma^{\mu}\partial_\mu\,-\,{\cal S}\,-\,\gamma^{\mu}\left( V_\mu\,+\,\gamma_5\, A_\mu\right)\right]q
\,-\,\frac{1}{4\,G_1}\,tr_f\left({\cal S}^2\,-\,m\,{\cal S}\,(U+U^\dagger)\right)\nonumber\\
& & \,+\,\frac{1}{4\,G_2}\,tr_f\left(({V}^\mu +{\cal V}^\mu_c -{\cal V}^\mu_\xi)^2\,+\,
({A}^\mu +{\cal A}^\mu_c -{\cal A}^\mu_\xi)^2\right).\label{LNJLB}
\end{eqnarray}
${\mathcal{V}}^\mu_c = \frac{i}{2}\left(\xi\partial^\mu\xi^\dagger\,+\,\xi^\dagger\partial^\mu\xi\right)$ and 
${\mathcal{A}}^\mu_c=\frac{i}{2}\left(\xi\partial^\mu\xi^\dagger\,-\,\xi^\dagger\partial^\mu\xi\right)$ are composite vector and axial vector fields. As apparent in Eq. (\ref{LNJLB}),  the coupling to  left (${\cal L}^\mu$) and right 
(${\cal R}^\mu$) electroweak currents is included through the replacement~: ${\cal V}^\mu_c\to{\cal V}^\mu_c -  {\cal V}^\mu_\xi$ and ${\cal A}^\mu_c\to{\cal A}^\mu_c -  {\cal A}^\mu_\xi$, with 
${\cal V}^\mu_\xi = \frac{1}{2}\left(\xi{\cal R}^\mu\xi^\dagger\,+\,\xi^\dagger{\cal L}^\mu\xi\right)$ and
${\cal A}^\mu_\xi = \frac{1}{2}\left(\xi{\cal R}^\mu\xi^\dagger\,-\,\xi^\dagger{\cal L}^\mu\xi\right)$.
The vacuum expectation value of the chiral invariant scalar field $\mathcal{S}$ coincides with the constituent quark mass $M_0\sim 350\,\mathrm{MeV}$. We define an "effective" or "nuclear physics" scalar field $S$ by rescaling the chiral invariant scalar field $\mathcal{S}$, according to :
\begin{equation}
 \mathcal{S}=\frac{M_0}{F_\pi}\,S=\frac{M_0}{F_\pi}\left(s+F_\pi\right).   
\end{equation}
Its vacuum expectation value coincides by construction with $F_\pi$, the pion decay constant obtained below after the full bosonization procedure. Its fluctuating piece, i.e., the  $s$ field, has to be identified with the usual "nuclear physics sigma meson" of relativistic Walecka theories, $\sigma_W$. In the simplest NJL picture this "nuclear physics sigma field" coincides with the canonical scalar field, $s_c$. However in presence of $\pi-a_1$ mixing (i.e. for a NJL model with a vector interaction) and from some details of the formal construction of the scalar field, the two fields are related by $s=z_S\,s_c$. The explicit form of the dimensionless normalization factor factor, $z_S$, in general slightly larger than unity, will be given below in terms of NJL loop integrals. This normalization factor has no influence in nuclear matter calculation at the Hartree level since the ratio of the scalar-nucleon coupling constant to the scalar (sigma) mass, $g_S/M_\sigma$, is not affected by this rescaling. However it may have a small influence on Fock terms.\\

As described in \cite{Chanfray2011}, the next step is to integrate out quarks in the Dirac sea, generating the quark fermion determinant; this constitutes the bosonization. In that way the kinetic energy of the meson fields will be dynamically generated from the quark loops, i.e., from quantum fluctuations. Then we perform a derivative expansion of this fermion determinant to second order in the derivatives. The difficulty lies in the fact that we do not make an expansion around a constant (vacuum expectation value of the scalar field) but we want to have a formal expansion with the scalar object $\mathcal{S}$ (or $s$) keeping its field status, so as to include its possible modification in the nuclear environment. For that purpose we use the elegant method proposed by Chan \cite{Chan}. This procedure is described in full details in the section 1 of our pervious paper \cite{Chanfray2011} where the explicit form of the mesonic lagrangian is given. It follows that all  the coupling constants or parameters will explicitly depends on the scalar field  $\mathcal{S}$ and then on density. We introduce various NJL loop integrals appearing in the bosonization procedure:
\begin{eqnarray}
&& I_0(\mathcal{S})=\int_0^\Lambda \frac{d{\bf p}}{(2\pi)^3}\,E_p(\mathcal{S}),\quad
I_1(\mathcal{S})=\int_0^\Lambda \frac{d{\bf p}}{(2\pi)^3}\,\frac{1}{2\,E_p(\mathcal{S})},\nonumber\\
&& I_2(\mathcal{S})=\int_0^\Lambda \frac{d{\bf p}}{(2\pi)^3}\,\frac{1}{4\,E^3_p(\mathcal{S})},\quad
 J_3(\mathcal{S})=\int_0^\Lambda\frac{d{\bf p}}{(2\pi)^3}\,\frac{3}{8\,E^5_p(\mathcal{S})}, \nonumber\\
&& J_4(\mathcal{S})=\int_0^\Lambda\frac{d{\bf p}}{(2\pi)^3}\,\frac{15}{16\,E^7_p(\mathcal{S})},\qquad\hbox{with}\qquad E_p=\sqrt{\mathcal{S}^2\,+\,p^2}\label{PARAM1}.
\end{eqnarray}
\subsection{The chiral effective potential}
The chiral effective potential can be obtained from the already mentioned bosonization. With vacuum value subtracted it reads:
\begin{equation}
V_{\chi,NJL}(s)=-2N_c N_f\,\big(I_0(\mathcal{S})\,-\,I_0(M_0)\big) \,+\,\frac{\left(\mathcal{S}-m\right)^2 -\left(M_0 - m\right)^2}{2\,G_1}.\label{VNJL}
\end{equation}
The quantity, $-2N_c N_f\,I_0(\mathcal{S})$, is nothing but the total energy of the Dirac sea of constituent quarks and $I_0(\mathcal{S})$ is a NJL loop integral given above in Eq. (\ref{PARAM1}).  The vacuum constituent quark mass $M_0$  corresponds to the minimum of the effective potential, i.e., $V'_{\chi,NJL}(s=0)=0$. It is consequently the solution of the gap equation:
\begin{equation}
M_0 = m\,+\,4N_c N_f M_0\,G_1\,I_1(M_0).    \label{GAP}
\end{equation}
 \\
In \cite{Universe, Chanfray2023} we have shown  (see Fig. 1 of Ref. \cite{Chanfray2023}) that the chiral effective potential can be well approximated, at not too large value of $s/F_\pi$ (i.e. at not too large value of the density), by its expansion to third order in $s$:
\begin{equation}
V_{\chi,\mathrm{NJL}}(s)= \frac{1}{2}\,M^2_\sigma\, {s}^2\, +\,\frac{1}{2}\,\frac{M^2_\sigma -M^2_\pi}{ F_\pi}\, {s}^3\,\big(1\,-\,C_{\chi,\mathrm{NJL}}\big) +...\, .\label{eq:vchiNJL2}. 
\end{equation}
where $F_\pi$ is the pion decay constant, $M_\pi$  the canonical pion mass and $M_\sigma$ the effective sigma mass obtained in the bosonized NJL model (see below). The specific NJL parameter $C_{\chi,\mathrm{NJL}}$ has been given in \cite{Chanfray2023,Universe}:
\begin{equation}
C_{\chi,\mathrm{NJL}}=\frac{2}{3}\,\frac{M^2_0\,J_3(M_0)}{I_2(M_0)}.    
\end{equation}
%%%%%%%%%%%%%%%%%%%%%%%%%%%%%%%%%%%%%%%%%%%%%%%%%%%%%%%%%%%%%%
\subsection{Pion mass, pion decay constant and effective sigma mass}
The four basic  parameters of the NJL model   are the coupling constants $G_1,_,G_2$, the cutoff $\Lambda$ and the current quark mass $m$. If the vector interaction is omitted ($G_2=0$), the pion decay constant , $f_\pi$, the pion mass, $m_\pi$, and a the effective sigma mass, $m_\sigma$, are such that:
\begin{equation}
 f^2_\pi=2N_c N_f M^2_0\,I_2(M_0), \qquad
\frac{m^2_\pi}{m}=\frac{M_0}{G_1\,f^2_\pi},\qquad m^2_\sigma=4\,M_0^2\,+\,m^2_\pi. \label{NJLpar}
\end{equation}
If the $\pi-a_1$ mixing is incorporated, through the introduction of a vector interaction with strength $G_2$ (see Eq. (1)
of Ref. \cite{Chanfray2011} or Eqs. (\ref{LNJL}, \ref{LNJLB}) above), the  pion decay constant parameter, the pion  mass parameter and the sigma mass parameter are modified according to: 
\begin{eqnarray}
&& f_\pi^2\to F_\pi^2=\frac{{f^2_\pi}}{1\,+\,4\,G_2\,f_\pi^2}, \qquad m_\pi^2\to M_\pi^2=\frac{M_0\, m}{G_1\,F^2_\pi}=m_\pi^2\,\left(1\,+\,4\,G_2\,f_\pi^2\right)\nonumber\\
&& m_\sigma \to M_\sigma=\frac{f_\pi}{F_\pi}\,m_\sigma=\sqrt{1\,+\,4\,G_2\,f_\pi^2}\,m_\sigma.
\label{G2}
\end{eqnarray}
$M_\pi$ is the canonical pion mass appearing in the bosonized NJL model (see for instance Eqs. (32,34) of \cite{Chanfray2011}) at the tree level,  i.e., ignoring vacuum pion loops fluctuations. It does not exactly coincide with the pion pole mass, $M_{\pi P}$,  of the full original NJL model (see Eq. (5) of Ref. \cite{Chanfray2011}). It is simple matter to show that:
\begin{equation}
 M_{\pi P}\simeq  M_\pi \left(1-\frac{m^2_\pi}{8 M_0^2}\,C_\chi\right)\quad\hbox{with}\qquad
 C_{\chi, NJL}=\frac{2}{3}\,\frac{M^2_0\,J_3(M_0)}{I_2(M_0)}\simeq 0.5.\label{PHYSMASS}
\end{equation}
With standard values of the NJL parameters used in this paper, the overestimation of the pion mass is indeed very small, not larger than $1 \% $. It is important to note that this pion mass, interpreted as the physical mass of the bosonized model, even if very close to the pole mass, has to be interpreted as a screening mass, related to the inverse of the genuine axial correlator at zero momentum. 
\\

We will see below, by direct inspection of the pion field coupling  to the weak current that, in the framework of the bosonized model at the tree level, the $F_\pi$ parameter defined by Eqs.~(\ref{NJLpar},\ref{G2}) has to be identified with the pion decay constant. However this above $F_\pi$ parameter {\it a priori} differs from the  value, $F_{\pi P}$, of the original full NJL model, by the residue $r_\pi$ at the pion pole (see the discussion at the beginning of \cite{Chanfray2011}), the value $r_\pi=1$  being strictly valid in the chiral limit. In effect, $F_{\pi P}$ is given by:
\begin{eqnarray}
 F^2_{\pi P}&=& M_0^2 \,2 N_c N_f\,\tilde{I}_2(M_{\pi P})\, r_\pi\qquad\hbox{with}\qquad
 \tilde{I}_2(\omega)=\frac{I_2(\omega)}{1+8\, M_0^2\, N_c N_f \,G_2\,I_2(\omega)},\nonumber\\
 I_2(\omega)&=&\int_0^\Lambda \frac{d{\bf p}}{(2\pi)^3}\,\frac{1}{E_p(M_0)\,(4\,E^2_p(M_0)-\omega^2)}, \quad r_\pi=\left[1+\frac{M^2_{\pi P}}{\tilde{I}_2(M_{\pi P})}\left(\frac{\partial \tilde{I}_2 }{\partial \omega^2}\right)_{\omega=M_{\pi P}}\right]^{-1}.
\end{eqnarray}
Expending $\tilde{I}_2(M_{\pi P})$ around $\omega=M_{\pi P}=0$, one obtains :
\begin{equation}
 F^2_{\pi P}\simeq M_0^2 \,2 N_c N_f\,\left[\tilde{I}_2+M^2_{\pi P}\left(\frac{\partial \tilde{I}_2 }{\partial \omega^2}\right)_{\omega=0}\right]\,  \left[1+\frac{M^2_{\pi P}}{\tilde{I}_2(M_{\pi P})}\left(\frac{\partial \tilde{I}_2 }{\partial \omega^2}\right)_{\omega=M_{\pi P}}\right]^{-1}\,\simeq F^2_\pi .
\end{equation}
We conclude that the two definitions of the pion decay constant coincide up to order $M^4_\pi$. Hence the bosonization procedure preserves the coupling to the electroweak sector.\\

From the form of the scalar field  kinetic energy term, $\mathcal{L}_{KS}=2\,N_c \,N_f \,I_{2S}(\mathcal{S})\,\partial^\mu\mathcal{S}\partial_\mu\mathcal{S}$ (see Eq. (20) of Ref. \cite{Chanfray2011}),  the $z_S$ parameter fixing the scale of the canonical scalar field (i.e., $s=z_S\,s_c$) and of the associated scalar "sigma meson"  is such that
\begin{equation}
 \frac{M_0}{F_\pi}\,z_s \equiv \left(\frac{1}{2\,N_c\, N_f \,I_{2S}(M_0)}\right)^{1/2}= \frac{M_0}{f_\pi}\,\left(\frac{I_2(M_0)}{I_{2S}(M_0)}\right)^{1/2},
\end{equation}
with $I_{2S}(\mathcal{S})=I_2(\mathcal{S})\,-\,\mathcal{S}^2\,J_3(\mathcal{S})\,+\,\frac{1}{9}\,\mathcal{S}^4\,J_4(\mathcal{S})$. Hence: 
\begin{equation}
z_s = z_A\,\left(\frac{I_2(M_0)}{I_{2S}(M_0)}\right)^{1/2} \qquad \hbox{with}\quad z_A=\frac{F_\pi}{f_\pi}=\frac{1}{\sqrt{1\,+\,4\,G_2\,f_\pi^2}}.\label{CANON}
\end{equation}
Contrary to the usual methods, the NJL loop integral, $I_{2S}$, appearing in the scalar kinetic energy term differs from the usual $I_2$ integral. In practice there is a significant compensation between $F_\pi/f_\pi <1$ and $\sqrt{I_2/I_{2S}}>1$, and $z_S$
never deviates from unity by more than $20\%$. Notice however that the difference between $I_2$ and $I_{2S}$ plays no role in the nuclear matter calculation in the Hartree approximation. It is also important to keep in mind that the derivation of Ref. \cite{Chanfray2011} has been done using implicitly a covariant regularization procedure and may be not strictly valid for a non covariant cutoff. A conservative attitude might be to maintain $I_{2S}=I_2$. However, both the effective sigma mass, $M_\sigma$, and the canonical sigma mass, $M_{\sigma c}$,
\begin{equation}
 M_\sigma=\frac{f_\pi}{F_\pi}\,m_\sigma ,\quad M_{\sigma c}=\frac{f_\pi}{F_\pi}\,z_S\,m_\sigma
 =\left(\frac{I_2(M_0)}{I_{2S}(M_0)}\right)^{1/2}\,m_\sigma ,
\end{equation}
deviate from the usual NJL sigma meson mass: $m_\sigma \sim 2 M_0$. Starting with typical constituent quark mass $M_0\sim 350 \,\mathrm{MeV}$, the usual NJL sigma mass, $m_\sigma \sim 700\,\mathrm{MeV}$, is rather low but including this scale correction, brings the canonical sigma mass closer to the linear sigma model value, as used in our previous works.
In reality even the canonical sigma mass of the bosonized NJL model does not coincide with the pole mass calculated in the original full NJL model and may even be significantly different. This quantity, $M_\sigma$, which appears in the bosonization procedure  has to be interpreted as a screening mass, related to the inverse of the scalar propagator at zero momentum.  \\
Also notice that the ratio, $g_S/M_\sigma=g_{S c}/M_{\sigma c}$, is not affected when passing  from the field $s$ to the the canonical scalar field $s_c$ and this rescaling has no effect at the Hartree level.\\

\subsection{Chiral limit}
If we introduce the constituent quark mass, $M_{0\chi L}$, taken in the limit of vanishing current quark mass as the solution of the gap equation: $1 = 4N_c N_f \,G_1\,I_1(M_{0\chi L})$ for a given set of NJL parameter ($G_1, m, \Lambda$),  the pion decay constant in the chiral limit is
\begin{equation}
F^2_\pi(M_{0\chi L})=F^2_\pi(m=0)=F^2,
\end{equation}
$F$ being the low energy constant appearing in leading order of chiral perturbation theory. One can also introduce the low energy constant, $B$, defined from the quark condensate in the chiral limit $B=-<\bar{q}\, q>_{\chi L}/F^2$. It is given in the model by $B=M_{0\chi L}/2 G_1 F^2$.
It follows that, to leading order in the current quark mass, the pion mass is given by the Gell-Mann-Oakes-Renner relation:
\begin{equation}
\mathcal{M}^2_\pi= -\frac{2 m \,<\bar{q}\,q>_{\chi L}}{F^2}=\frac{m \,M_{0\chi L}}{G_1\,F^2}.   
\end{equation}
Using the explicit NJL dependence 
of the constituent quark mass upon the current quark mass,
\begin{equation}
m\,\frac{\partial M_0}{\partial m}=M_0 \,\frac{m^2_\pi}{ m^2_\sigma}=M_0 \,\frac{M^2_\pi}{ M^2_\sigma},   
\end{equation}
one obtains an expansion of the pion mass and of the pion decay constant in the quark mass (or equivalently $\mathcal{M}^2_\pi$ which is proportional to the quark mass). It is thus  a simple matter  to show that to second order in the quark mass, one has
\begin{eqnarray}
F^2_\pi &=& F^2\left[1\,+\,\frac{\mathcal{M}^2_\pi}{2M^2_{0\chi L}}\left(1-\frac{3}{2}C^0_\chi\right)\,z^4_A\right]\nonumber\\
M^2_\pi  &=&\mathcal{M}^2_\pi\left[1\,-\,\frac{\mathcal{M}^2_\pi}{2M^2_{0\chi L}}\left(\left(1-\frac{3}{2}C^0_\chi\right)\,z^4_A-\frac{z^2_A}{2}\right)\right],
\qquad z_A=\frac{F_\pi}{f_\pi},\label{EXP}
\end{eqnarray}
where it is understood that the NJL parameters, $C^0_\chi,\,z_A$, are taken in the chiral limit, which are in practice extremely close to their values taken at the physical pion mass.

%%%%%%%%%%%%%%%%%%%%%%%%%%%%%%%%%%%%%%%%%%%%%%%%%%%%%%%%%%%%%%%%%%%%%%%
\subsection{The pionic sector and  and the pionic self-energy of the nucleon}
%%%%%%%%%%%%%%%%%%%%%%%%%%%%%%%%%%%%%%%
Once quarks in the Dirac sea have been integrated out, one remains with the pure meson sector lagrangian and the Yukawa-like  lagrangian describing the coupling of the vector, axial-vector and scalar fields to the valence quarks (see Eq. (28) of Ref. \cite{Chanfray2011}) :
\begin{equation}
  {\cal L}_{val}= \bar{q_v}\left[i\,\gamma^{\mu}\partial_\mu\,-\,{\cal S}\,-\,\gamma^{\mu}\left( V_\mu\,+\,\gamma_5\, A_\mu\right)\right]q_v . 
\end{equation}
In the following we will note most of the time this valence quark field $q$ and not $q_v$ for simplicity. The vector field $V^\mu$ is expressed in terms of the canonical vector field $v^\mu$ \footnote{Rigorously the canonical vector field is ${v_\mu}_c= [g_V(\mathcal{S})/g_V(\mathcal{S}=M_0)]\,v_\mu$} and a composite vector field $\mathcal{V}_c$ according to 
$$
 V_\mu=  - \mathcal{V}_c^{\mu}\, +g_V(\mathcal{S})\,v_\mu,
$$
where $g_V(\mathcal{S})$, whose explicit expression is not needed in the present paper, is the quark-vector coupling constant given by Eq. (30) of Ref \cite{Chanfray2011}.
We also redefine the axial-vector field $A_\mu$ in order to eliminate the $\pi-a_1$ mixing according to \cite{Chanfray2011}:
$$
 A_\mu=  - \frac{\mathcal{A}_c^{\mu}}{1\,+\,4\,G_2\,{\cal S}^2\,I({\cal S})}_,+g_V(\mathcal{S})\,a_\mu.
$$
$a_\mu$ is the residual canonical axial-vector field with a large (screening) mass (typically larger than $1.2\,GeV$, see \cite{Chanfray2011})  such that $g^2_V/M^2_A=G_2/[1\,+\,4\,G_2\,{\cal S}^2\,I({\cal S})]$. $\mathcal{A}_c^{\mu}$ is a composite pseudo-vector field having a pionic content. Omitting the vector meson and axial-vector meson mass and kinetic terms, it follows that the the scalar-pion-valence quark lagrangian reads:
\begin{eqnarray}\
		{\cal L}_{q\pi s}&=&\bar{q}_v\left[	i\gamma_{\mu}\left(\partial^{\mu} - i \mathcal{V}_c^{\mu}+ ig_V\, v_\mu +i g_V\, a_\mu \gamma_5\right) +z^2_A({\cal S}) \gamma_{\mu}\gamma^{5} \mathcal{A}_c^{\mu} - \mathcal{S}\right]q_{v}\,\nonumber\\
  &&+\,
		\frac{1}{2}\,2 N_c N_f\,I_{2S}({\cal S})\,\partial^{\mu}{\cal S}\,\partial_{\mu}{\cal S}\,- \,V_\chi({\cal S})\nonumber\\
& & +\,\frac{m\,{\cal S}}{4\,G_1}\,tr_f(U\, +\, U^\dagger\,-\,2)\,+\,
\tilde I({\cal S})\,{\cal S}^2\,tr_f\left({\cal A}^\mu_c\cdot{\cal A}^c_\mu\right),\nonumber\\
		\hbox{with} && U(x)=e^{i\,\vec{\tau}\cdot\vec{\Phi}(x)/F_\Phi}, \qquad\xi=\sqrt{U},\nonumber\\
  &&\mathcal{V}_c^{\mu}=\frac{i}{2}\left(\xi\partial^{\mu}\xi^\dagger + \xi^\dagger\partial^{\mu}\xi\right),\qquad 
		\mathcal{A}_c^{\mu}=\frac{i}{2}\left(\xi\partial^{\mu}\xi^\dagger - \xi^\dagger\partial^{\mu}\xi\right), \nonumber\\
		&&I({\cal S})=2 N_c N_f\, I_{2}({\cal S}),\qquad
\tilde I({\cal S})\equiv 2 N_c N_f\,\tilde I_{2}({\cal S})=\frac{I({\cal S})}{1\,+\,4\,G_2\,\bar{\cal S}^2\,I({\cal S})},\nonumber\\
&&  f^2_\pi({\cal S})={\cal S}^2 I_{2}({\cal S}),
\qquad F^2_\pi({\cal S})={\cal S}^2 \tilde I_{2}({\cal S}),\nonumber\\
&&z_A({\cal S})=\frac{F_\pi({\cal S})}{f_\pi({\cal S})}=\frac{1}{\sqrt{1\,+\,4\,G_2\,{\cal S}^2\,I({\cal S})}}.\label{LQPI}
	\end{eqnarray}
The canonical pion field, $\Phi\equiv\vec\tau\cdot\vec\Phi$, is defined  through $U=exp(i\Phi/F_\Phi)$ where the constant $F_\Phi$ is given by  $F_\Phi^2=\tilde I(M_0)\,M^{2}_{0}\equiv F^2_\pi$. Hence to leading order in the pion field one has $\mathcal{A}_c^{\mu}= \partial^\mu \Phi/F_\pi+ \mathcal{O}(\Phi^2/F_\pi^2)$. We can remark on the first line of Eq. (\ref{LQPI}) that the pseudo-vector derivative coupling  $\gamma_\mu\gamma_5 \partial^\mu \Phi$ is apparently reduced by a factor $z^2_A({\cal S})=F^2_\pi({\cal S})/f^2_\pi({\cal S})$ when $G_2$ is non zero. 
However the presence of the $g_V\,\bar{q}\gamma^\mu\gamma_5\, a_\mu q$ piece in the lagrangian induces a vertex correction to the $\gamma_\mu\gamma_5$ coupling. This correction involves the emission at the pseudo-vector (PV) vertex of a heavy axial-vector field which subsequently couples  to a $q\bar{q}$ polarization bubble ending with a PV $\gamma_\mu\gamma_5$ coupling, i.e.:
$$
g_V\,\left(\frac{-1}{M^2_A}\right)\,g_V\,\big(-4\mathcal{S}^2\,I(\mathcal{S})\big)\,\gamma^\mu\gamma_5=\frac{G_2\,4\mathcal{S}^2\,I(\mathcal{S})}{1\,+\,4\,G_2\,{\cal S}^2\,I({\cal S})}\gamma^\mu\gamma_5=\big(1-z_A^2(\mathcal{S})\big)\gamma^\mu\gamma_5.
$$
Hence the net effect of the coupling to the residual heavy axial-vector field is simply to eliminate the $z^2_A(\mathcal{S})$ factor in front of  the pion-valence quark PV-coupling lagrangian.\\

The coupling to electroweak current is obtained through the replacement ${\cal A}^\mu_c\to{\cal A}^\mu_c -  {\cal A}^\mu_\xi$ in the meson sector lagrangian. The relevant piece is: 
\begin{equation}
 \mathcal{L}_W=  -2\,\tilde I({\cal S})\,{\cal S}^2\,tr_f  ({\cal A}^\mu_c\cdot{\cal A}^\mu_\xi)\,-\,\frac{g_V}{2G_2}\,tr_f (a_\mu\cdot{\cal A}^\mu_\xi).
\end{equation}
From the equation of motion, the $a_\mu$ field induced by the presence of the valence quark is such that:
\begin{equation}
 \frac{g_V}{2G_2}\,a_\mu =\frac{1}{G_2}\,\frac{g^2_V}{M^2_A} \,\bar{q} \gamma_{\mu}\gamma^{5}q=\frac{1}{1\,+\,4\,G_2\,{\cal S}^2\,I({\cal S})}\bar{q} \gamma_{\mu}\gamma^{5}q=z^2_A(\mathcal{S})\,\bar{q} \gamma_{\mu}\gamma^{5}q.
\end{equation}
Again the $\gamma_{\mu}\gamma^{5}$ vertex correction will just cancel the $z^2_A(\mathcal{S})$ factor in front of the quark axial current. This finally generates the electroweak lagrangian as:
\begin{equation}
\mathcal{L}_W= -\tilde I({\cal S})\,{\cal S}^2\,tr_f\left(\left(\xi {\cal A}^\mu_c \xi^\dagger +\xi^\dagger {\cal A}^\mu_c\xi\right) \frac{\vec{\tau}\cdot\vec{W}}{2}\right)\,-\,
\bar{q} \gamma_{\mu}\gamma^{5}\,\frac{ \xi\,\vec{\tau}\cdot\vec{W}\,\xi^\dagger + \xi^\dagger\,\vec{\tau}\cdot\vec{W}\,\xi}{2}\,q.\label{WeakL}
\end{equation}
A direct inspection of the coupling of $\partial^\mu\Phi$ to the axial weak current allows to confirm that  the tree level pion decay constant has to be identified with the parameter $F_\pi$.\\

To leading order in the pion field, one gets an effective lagrangian involving the dominant p-wave coupling of the pion to the valence quarks:
\begin{eqnarray}
{\cal L}_{q\pi s}&=&\bar{q}_v\left[i\gamma_{\mu}\partial^{\mu}  - \mathcal{S} -\frac{1}{F_\pi}\gamma^j\partial_j\vec{\Phi}\cdot\vec{\tau} \right]q_{v}\,-\,V_\chi({\cal S})\nonumber\\
&&+\frac{1}{2} \frac{Z_s(\mathcal{S})}{z^2_S} \,\partial^{\mu} s\,\partial_{\mu} s\,-\,\frac{1}{2}\mu_\pi^2(\mathcal{S})\,\vec{\Phi}^2 \,+\, \frac{1}{2} Z_\pi(\mathcal{S}) \,\partial^{\mu}\vec{\Phi}\cdot\partial_{\mu}\vec{\Phi}\nonumber\\
\hbox{with}&&\quad Z_s(\mathcal{S})=\frac{I_{2S}(\mathcal{S})}{I_{2S}(M_0)},\quad Z_\pi(\mathcal{S})=\frac{\mathcal{S}^2\tilde{I}_2(\mathcal{S})}{M_0^2\tilde{I}_2(M_0)} ,\quad \mu_\pi^2(\mathcal{S})=\frac{\mathcal{S}}{M_0}\,M^2_\pi.\label{LQPI1}
\end{eqnarray}
A standard calculation thus allows to obtain the pionic self-energy of the in-medium nucleon (see Eq. (22) of Ref. \cite{Chanfray2007}),
\begin{equation}
\Sigma^{(\pi)}({\mathcal{S};m})=-{3\over 2}\left({g_A(\mathcal{S})\over 2 F_\pi(\mathcal{S})}\right)^2\int{d{\bf q}\over 
(2\pi)^3} {\bf q}^2 \,v^2({\bf q}; \mathcal{S})\left(\frac{1}{\omega_q}\frac{1}{\omega_q+\epsilon_{N {\bf q}}}
+\frac{32}{25}\,\frac{1}{\omega_q}\frac{1}{\omega_q+\epsilon_{\Delta{\bf
q}}}\right),\label{SELFENERGY}
\end{equation}
which depends on the in-medium pion decay constant and on the in-medium pion mass: 
\begin{equation}
F^2_\pi(\mathcal{S})=\mathcal{S}^2\,\Tilde{I}_2 (\mathcal{S}),\quad \omega_q=\sqrt{q^2+M^2_\pi(\mathcal{S})},\quad
M^2_\pi(\mathcal{S})=\frac{\mathcal{S}}{M_0}\frac{F^2_\pi}{F^2_\pi(\mathcal{S})}\,M^2_\pi\,=\,\frac{m \mathcal{S}}{G_1 F^2_\pi(\mathcal{S})}.
\end{equation}
Notice that, in the above expression of the pionic self-energy,  we have taken into account recoil corrections with $\epsilon_{N {\bf{q}}}={\bf{q}}^2/2 M_N$ and 
$\epsilon_{\Delta {\bf{q}}}=\omega_\Delta  +{\bf{q}}^2/2 M_\Delta$, $\omega_\Delta=M_\Delta -M_N$. The axial coupling constant, $g_A(\mathcal{S})$, and the form factor appearing in the expression of the self-energy also depend on the background scalar field ${\mathcal{S}}$, whose value coincides with the in-medium modified constituent quark mass.  This expression of the pionic-self-energy, as well as the quark core nucleon mass discussed in the next section,  should be seen as a functional of the scalar field $\mathcal{S}$. In the spirit of the Born-Oppenheimer approximation this value of $\mathcal{S}$ coincides with the value of the scalar field $\mathcal{S}(\bfvec{R}_N)=(M_0/F_\pi)(F_\pi + s(\bfvec{R}_N))$ taken at the nucleon center of mass (CM) position, $\bfvec{R}_N$, hence assuming that
 the variation of the scalar field  is slow enough to neglect its variation in the nucleonic volume, $V_N\sim 1/\sigma^{3/2}$. This reasoning  is presented in full in various QMC papers \cite{Guichon96,QMCreview}.

The combination $g_A(\mathcal{S})\,v({\bf q};\mathcal{S} )$   can be calculated in a given confining model as explicitly done in the next section. One  explicit expression (ignoring  finite pion size effects and center of mass corrections) is 
\begin{equation}
g_A(\mathcal{S})\,v({\bf q}; \mathcal{S})=\frac{5}{3}\,\int d^3r\,e^{i\bf{q}\cdot\bf{r}}\,\left(u^2(r;\mathcal{S})-\frac{1}{3}v^2(r; \mathcal{S})\right)\label{GAC},
\end{equation}
where $u(r;\mathcal{S})$ and $v(r;\mathcal{S})$ are the up and down component of the Dirac wave function of the core lowest energy orbital. Starting from the expression of the weak current lagrangian (Eq. \ref{WeakL}), one can check that $g_A(M_0)$ coincides with the nucleon weak axial coupling constant at the tree level. Hence the in-medium  $\pi NN$ coupling constant, $G_{\pi NN}(\mathcal{S})=g_A(\mathcal{S})/2F_\pi(\mathcal{S})$ coincides at zero density (rigorously in the chiral limit) with the coupling, $g_A/2F_\pi$, appearing in the leading term of baryon chiral perturbation theory.
\\

For completeness, let us mention that one very important quantity, very sensitive to the nucleon internal structure, is the nucleon sigma term which is defined as the matrix element of the piece of the hamiltonian breaking explicitly chiral symmetry with vacuum value subtracted:
\begin{eqnarray}
\sigma_N&=&\langle N |H_{\chi SB}|N\rangle , \nonumber\\
H_{\chi SB}&=&\int d^3r\frac{m}{4G_1}\,\mathcal{S}({\bf{r}})\,tr_f \left(U+U^\dagger\right)({\bf{r}})=\int d^3r\frac{F^2_\pi M^2_\pi}{4 M_0}\,\mathcal{S}({\bf{r}})\,tr_f \left(U+U^\dagger\right)({\bf{r}}).
\end{eqnarray}
It can be obtained as the expectation value of $H_{\chi SB}$ in the limit of vanishing density:
\begin{equation}
\sigma_N= \int d^3r \, F_\pi\, M^2_\pi  \,\bar{s}\, +\,  \int d^3r\,\frac{1}{2} M^2_\pi\,\Phi^2({\bf{r}})\equiv
\sigma_N^{(s)}\,+\,\sigma_N^{(\pi)}.
\end{equation}
The non pionic part can be obtained from the gap equation in the limit of zero density (see, e.g, section III of Ref. \cite{Chanfray2007})
\begin{equation}
\sigma_N^{(s)}={F_\pi}\, g_S\,\frac{M^2_\pi}{M^2_\sigma}= F_\pi\, g_S\,\frac{m^2_\pi}{m^2_\sigma}\simeq 21\,\mathrm{MeV},
\end{equation}
where the numerical value is obtained either with parameter set NJLset1 or NJLset2.
The pionic piece can be obtained directly from the pion propagator as in \cite{Chanfray2007}:
\begin{eqnarray}
\sigma_N^{(\pi)}&=&{3\over 2}\left({g_A(M_0)\over 2 F_\pi(M_0)}\right)^2 M^2_\pi\int{d{\bf q}\over 
(2\pi)^3} \frac{{\bf q}^2 \,v^2({\bf q})}{2 \omega^2_q}\bigg[\frac{1}{\omega_q}\frac{1}{\omega_q+\epsilon_{N {\bf q}}}
+\frac{1}{(\omega_q+\epsilon_{N {\bf q}})^2}\nonumber\\
&&+\frac{32}{25}\,\bigg(\frac{1}{\omega_q}\frac{1}{\omega_q+\epsilon_{\Delta{\bf
q}}} + \frac{1}{(\omega_q+\epsilon_{\Delta{\bf
q}})^2}\bigg)\bigg] .
\end{eqnarray}
One can notice that one formally has: 
\begin{equation}
\sigma_N^{(\pi)}=M^2_\pi\,\frac{d\Sigma^{(\pi)}(M_0 ;m)}{d M^2_\pi}\equiv m\,\frac{d\Sigma^{(\pi)}(M_0 ;m)}{d m}
\quad\hbox{with}\quad M^2_\pi=\frac{M_0}{G_1 F^2_\pi}\,m,
\end{equation}
where the derivative is formally taken with respect to the pion mass appearing in the energy denominators, ignoring the implicit possible quark mass dependence of $M^2_\pi/m$. This light quark sigma term has been abundantly discussed in our previous papers \cite{Chanfray2007,Massot2008,Chanfray2011}. \\
%%%%%%%%%%%%%%%%%%%%%%%%%%%%%%%%%%%%%%%%%%%%%
\subsection{Vacuum pion loops correction and connection with chiral perturbation theory}
%%%%%%%%%%%%%%%%%%%%%%%%%%%%%%%%%%%%%%%%%%%%%
The third line of the lagrangian (Eq. \ref{LQPI}) actually generates a four-pion lagrangian which has been ignored in  
Eq. (\ref{LQPI1}) and whose explicit form is given in Ref. \cite{Chanfray1999}. As in this reference we temporarily represent the unitary matrix as $U=exp\left(\vec{\tau}\cdot\hat{\Phi}(x)\mathcal{F}(X)/F_\pi\right)$ with $X^2=\Phi^2/F^2_\pi$ and $\mathcal{F}(X)=X+\alpha X^3 +\alpha' X^5 +...$; we also introduce the particular combination $\beta=1+10(\alpha -1/6)$. Since the physical observables should not depend on the representation, the independence against $\alpha$ will provide a consistency check of the results. This four-pion lagrangian, formally similar to the non linear sigma model lagrangian, modifies the pion propagator according to \cite{Chanfray1999}:
\begin{equation}
D(k,\omega)=\frac{Z_R}{\omega^2-k^2-M^2_\pi\left(1\,+\,\frac{1}{6}\,\frac{<\Phi^2>}{F^2_\pi}\right)}\quad\hbox{with}\quad Z_R=\left(1\,+\,\frac{\beta}{3}\,\frac{<\Phi^2>}{F^2_\pi}\right)^{-1}.
\end{equation}
Hence  this loop correction depending on the vacuum pion scalar density $<\Phi^2>$, often referred as associated with a tadpole diagram,  induces, as at finite temperature, the following change of the pion mass:
\begin{equation}
 M^2_{\pi P}\quad\to\quad M^2_{\pi L}=M^2_{\pi P}\,\left(1\,+\,\frac{1}{6}\,\frac{<\Phi^2>}{F^2_\pi}\right).\label{MPIL}
\end{equation}
This loop correction also induces a renormalization of the weak axial current which can be deduced from the pionic piece of the weak lagrangian (Eq. \ref{WeakL}):
\begin{equation}
\vec{\mathcal{A}}^\mu_W= F_\pi\left[\partial^\mu\vec{\Phi}\left(1+\left(\alpha-\frac{2}{3}\right)\frac{\Phi^2}{F^2_\pi} \right)+\left(2\,\alpha+\frac{2}{3} \right)\vec{\Phi}\,\vec{\Phi}\cdot\partial^\mu\vec{\Phi}\right].   
\end{equation}
It follows that the pion decay constant is modified, to one loop order, according to:
\begin{equation}
 F^2_{\pi}\quad\to\quad F^2_{\pi L}=F^2_\pi\,{Z_R}\, \,\left(1\,+\,2\,\left(\frac{\beta}{6}-\frac{1}{3}\right)\frac{<\Phi^2>}{F^2_\pi}\right)\,=\,
 F^2_\pi \,\left(1\,-\,\frac{4}{6}\,\frac{<\Phi^2>}{F^2_\pi}\right).\label{FPIL}
\end{equation}
The pion scalar density generates, after one constant subtraction, the so called chiral logarithm,
\begin{equation}
<\Phi^2( M^2_{\pi})> =3\int \frac{d^4 p}{(2\pi)^4}\frac{1}{p^2\,+\,M^2_{\pi}} =3\,\frac{M^2_{\pi}}{16\,\pi^2} \ln{\frac{M^2_{\pi}}{\Lambda_L^2}},
\end{equation}
where the cutoff $\Lambda_L$ is a new model parameter expected to be of the order of the NJL cutoff. To make connection with chiral perturbation theory ($\chi PT$), we now consider the expansion of the true physical quantities $M^2_{\pi L}$
and $F^2_{\pi L}$. 
Using Eqs. (\ref{PHYSMASS},\ref{EXP},\ref{MPIL},\ref{FPIL}) one obtains a NLO expansion of the pion mass and of the pion decay constant in the quark mass (or equivalently $\mathcal{M}^2_{\pi}$ which is proportional to the quark mass):
\begin{eqnarray}
 M^2_{\pi L} &=& \mathcal{M}^2_{\pi}\,\left[1\,-\,\frac{\mathcal{M}^2_{\pi}}{2M^2_{0\chi L}}\,(1\,-\,\frac{3}{2}\,C^0_\chi)\,z^4_A\,+\frac{\mathcal{M}^2_{\pi}}{4M^2_{0\chi L}}\left(1-C^0_\chi\right) z^2_A\,+\,\frac{\mathcal{M}^2_{\pi}}{32\,\pi^2\,F^2}\, \ln{\frac{\mathcal{M}^2_{\pi}}{\Lambda_L^2}}\right] \nonumber\\
 &\equiv& \mathcal{M}^2_{\pi}\,\left[1-\,\frac{\mathcal{M}^2_{\pi}}{32\,\pi^2\,F^2}\,\, \bar{l}_3\,+\,\mathcal{O}(\mathcal{M}^4_{\pi})\right],
\end{eqnarray}
\begin{eqnarray}
 F^2_{\pi L } &=& F^2 \,\left[1\,+\,\frac{2 \,\mathcal{M}^2_{\pi}}{4M^2_{0\chi L}}\,(1\,-\,\frac{3}{2}\,C^0_\chi)z^4_A\,-\,\frac{\mathcal{M}^2_{\pi}}{8\,\pi^2\,F^2}\, \ln{\frac{\mathcal{M}^2_{\pi}}{\Lambda_L^2}}\right] \nonumber\\
 &\equiv& F^2\,\left[1+\,\frac{\mathcal{M}^2_{\pi}}{8\,\pi^2\,F^2} \,\,\bar{l}_4\,+\,\mathcal{O}(\mathcal{M}^4_{\pi})\right].
\end{eqnarray}
In the above two equations, the right-hand-sides represent the  NLO results depending on to LEC's, $\bar{l}_3$ and $\bar{l}_4$, equivalently represented by two scale parameters, $\Lambda_3$ and $\Lambda_4$, according to: 
\begin{equation}
\bar{l}_3=\ln{\frac{\Lambda_3^2}{\mathcal{M}^2_{\pi}}}\,\qquad   \bar{l}_4=\ln{\frac{\Lambda_4^2}{\mathcal{M}^2_{\pi}}}. 
\end{equation}
In view of the remarkable progress achieved with simulations of QCD on a lattice, the dependence of the pion mass on the quark masses can now be calculated from first principles. The data indeed show evidence for the presence of the logarithmic term. The same is true for the pion decay constant.  The numerical results obtained in recent determinations yield  $\bar{l}_3=3.0\pm 0.3$, $\bar{l}_4=4.3\pm 0.3$. These results imply $\Lambda_3 =0.63 \pm 0.06_,GeV$ and $\Lambda_4 =1.12 \pm 0.12\,GeV$ \cite{Leut2012}. The NJL model value for these two LEC's are:
 \begin{eqnarray}
&&\Lambda_3=exp\left[\left(\left(1\,-\,\frac{3}{2}\,C^0_\chi\right)\,z^4_A-\frac{1}{2}\left(1\,-\,C^0_\chi\right)z^2_A\right)\,\frac{8\pi^2\,F^2}{M^2_{0\chi L}}\right]\,\Lambda_L,\nonumber\\
&&
\Lambda_4=exp\left[\left((1\,-\,\frac{3}{2}\,C^0_\chi\right)z^4_A\,\frac{2\pi^2\,F^2}{M^2_{0\chi L}}\right]\,\Lambda_L,\nonumber\\
&&\quad\to\quad
\frac{\Lambda_4}{\Lambda_3}=exp\left[\left( 3\,\left( \frac{3}{2}\,C^0_\chi\,-\,1\right)z^4_A\,+\,2\left(1\,-\,C^0_\chi\right)z^2_A\right)\,\frac{2\pi^2\,F^2}{M^2_{0\chi L}}\right].
 \end{eqnarray}
 Let us first consider the case of the NJL model ignoring the pion-axial mixing, i.e., $z_A=1$. Since $C^0_\chi\sim 0.5$, $\Lambda_3$ is very close to $\Lambda_L$. Hence it turns out that the compatibility with $\chi PT$ requires a pion loop cutoff very close to the original NJL cutoff around $0.6\,GeV$. What we  also see is that $\Lambda_4$ is larger than $\Lambda_3$ provided $C^0_\chi>0.4$ which is actually the case with our typical set of parameters ($M_0\sim 350\, \mathrm{MeV},\, \Lambda=600\,\mathrm{MeV}$). This can be seen as a  very encouraging point even if  the ratio predicted by the NJL model ($\Lambda_4/\Lambda_3\sim 1.4$ for $C^0_\chi\sim 0.5$)  is smaller than $1.8$, which is the value extracted from lattice QCD. In a version incorporating the pion-axial mixing  ($G_2=G_1$, $M_0\sim 360\,\mathrm{MeV},\,\Lambda\sim 740\,\mathrm{MeV}, \,C^0_\chi\sim 0.43, \,z_A=0.87$), the situation is even better  since one finds $\Lambda_4/\Lambda_3\sim 1.7$. 
 In addition it is worthwhile to mention that this fact, i.e., the compatibility of the NJL model with lattice $\chi PT$ results,    is due to the sizable value of the NJL loop parameter $C_\chi^0\sim 0.5$ entering the chiral effective potential and consequently the chiral susceptibility, $a_4$, and the effective repulsive three-body force contributing to the saturation mechanism! Conversely the lattice $\chi PT$ values for these LEC's can be viewed as a constraint on the parameter $C^0_\chi$ controlling one sizable part of  the repulsive three-body forces needed for the saturation mechanism to occur.\\
 Numerically this  loop correction for the pion mass is however very tenuous, less than $1\%$, and thus invisible in the numerical nuclear matter calculation. Nevertheless this logarithmic quark mass dependence may  alter the constraints provided by the chiral extrapolation of the nucleon mass on the model parameters of our approach. The effect on the pion decay constant is more significant ($F_\chi/F\simeq 1.04-1.06$ with our parameters to be compared with $1.07$ from lattice), yielding a pion decay constant in the chiral limit  $F\simeq 85-87\,\mathrm{MeV}$. Rigorously, the parameters of the NJL model should be fixed taking the chiral limit of the pion decay constant as a constraint on the model parameters.
  \\
  
 The pion loop contribution might also affect the $\pi NN$ coupling constant $G_{\pi NN}\sim g_A/F_\pi$. However it turns out that there is no renormalization to one-loop order as shown in \cite{Chanfray1999}. Hence in presence of pion loop correction, the pionic self-energy is modified only by the modification of the pion mass from the chiral logarithm. 
 %%%%%%%%%%%%%%%%%%%%%%%%%%
 \subsection{Pion loops correction at finite density and temperature}
 %%%%%%%%%%%%%%%%%%%%%%%
 At finite density and/or temperature, The in-medium pion mass  entering the expression of the pion-self-energy (Eq. \ref{SELFENERGY}), is modified according to \cite{Chanfray1999}:
 \begin{equation}
 M^2_{\pi}(M)\quad\to\quad M^2_{\pi }(M)\,\left(1\,+\,\frac{1}{6}\,\frac{<\Phi^2>(\rho)}{F^2_\pi}\right).\label{MPIMED}
\end{equation}
A complete expression of the pion scalar density at finite temperature and density in terms of the longitudinal spin-isospin response function can be found in our previous papers, see e.g. \cite{Chanfray1999b}. To leading order in density it reads 
 \begin{equation}
 <\Phi^2>(\rho)=2\rho\,\frac{\sigma^{(\pi)}_N}{M^2_\pi},
\end{equation}
where $\sigma^{(\pi)}_N$ is the pionic contribution to the nucleon sigma term. Hence the pion mass entering the pionic self-energy is modified approximately as:
\begin{equation}
 \left(\frac{\Delta M^2_\pi}{M^2_\pi} \right)_{pion\, loop}\simeq 0.06\,\frac{\rho}{\rho_0} \, \frac{\sigma^{(\pi)}_N}{25\,\mathrm{MeV}}. 
\end{equation}
In passing one can notice that, at finite temperature, the pion scalar density taken in the chiral limit is 
\begin{equation}
<\Phi^2>(T)= \int \frac{d^3 k}{(2\pi)^3} \,\frac{f_k (T)}{\omega_k} =\frac{T^2}{24},
\end{equation}
which allows to recover the finite temperature result for the pion mass and the pion decay constant obtained by Gasser and Leutwyler \cite{Gasser}.
\\

Again, to one loop order, there is no renormalization of the $\pi NN$ coupling constant \cite{Chanfray1999}. Hence the only modification of the pionic self-energy induced by pion loop at finite density is the loop correction to the pion mass as for the vacuum loop correction.\\

Before going to the explicit calculation of the confined quark wave function, let us summarize what we have achieved in this section: we started with the FCM inspired NJL lagrangian to generate the chiral scalar potential, the mesonic sector (in particular the pion field) and its coupling to the valence quarks whose mass coincides with the scalar field.  

\section{Confined quark wave function, in-medium nucleon mass and the response parameters to the nuclear scalar field }
\subsection{The "shifted vacuum"}
 The aim of this section is to establish the bound state equation satisfied by the confined  constituent quark submitted to the confined interaction $V_C$ and moving in the scalar field 
$\mathcal{S}$ considered as an external slowly varying field (hence taken as a uniform field within the nucleonic volume) in the spirit of the Born-Oppenheimer approximation (also look at the discussion after Eq. (\ref{SELFENERGY}) in section 4). To achieve this program we introduce a particular state, $\left|\varphi(\mathcal{S})\right\rangle$, (previously noted $\left|\varphi(\rho)\right\rangle$ since $\mathcal{S}$ is a function of the nucleonic density $\rho$) which is the vacuum of constituent quarks with mass $M=\mathcal{S}$, the true vacuum introduced in section 3 being just $\left|\varphi\right\rangle=\left|\varphi(\mathcal{S}=M_0)\right\rangle$.
This zero baryon number state is of BCS nature similar to the true vacuum but with a modified chiral phase:
\begin{equation}
\sin\varphi(p ;\mathcal{S})=\frac{{\mathcal{S}}}{E_p({\mathcal{S}})}\equiv\frac{{\mathcal{S}}}{\sqrt{p^2 +\bar{\mathcal{S}}^2}}.
\end{equation}
The deviation of the chiral phase with respect to the true vacuum value generates a decrease of the quark condensate
seen as a partial chiral symmetry restoration induced by the presence of the surrounding nucleons. By (slight) abuse of language, this zero baryon number state $\left|\varphi(\mathcal{S})\right\rangle$ is called a "shifted vacuum" but its precise definition is
\begin{eqnarray}
\left|\varphi(\mathcal{S})\right\rangle&=& \mathcal{N}\,exp\left(-\sum_{s,\, p<\Lambda}\,
\gamma_{ps}\,b^\dagger_{{\bf p} s}\,d^\dagger_{-{\bf p}-s}\right) \left|\Phi_0\right\rangle
= \prod_{s,\, p<\Lambda}\,\left(\alpha_p+s \,\beta_p
\,b^\dagger_{{\bf p} s}\,d^\dagger_{-{\bf p}-s}\right)\left|\Phi_0\right\rangle\nonumber\\
\hbox{with}&&\qquad\gamma_{ps}=-s\,\frac{\beta_p}{\alpha_p},\qquad \left(\begin{array}{c}
\alpha_p\\ \beta_p\end{array}\right)
=\bigg[\frac{1}{2}\left(1\pm \frac{p^2+\mathcal{S} m}{\epsilon_p E_p(\mathcal{S})}\right)\bigg]^{1/ 2},
\end{eqnarray}
where the state, $\left|\Phi_0\right\rangle$, is the perturbative vacuum, \textit{i.e.}, the vacuum of the bare quark operator $b_{{\bf p} s}$  with current mass $m$. In the above expressions all the parameters $\alpha_p, \beta_p, \gamma_p$ are actually $\mathcal{S}$ dependent parameters. The relationship between the bare and dressed operators is given by the Bogoliubov transformation,
$$b_{{\bf p} s}=\alpha_p(\mathcal{S}) B_{{\bf p} s}(\mathcal{S})\,+s\,\beta_p(\mathcal{S}) D^\dagger_{-{\bf p} -s}(\mathcal{S}),
\qquad
d^\dagger_{-{\bf p} -s}=-s\,\beta_p(\mathcal{S}) B_{{\bf p} s}(\mathcal{S})\,+\alpha_p(\mathcal{S}) 
D^\dagger_{-{\bf p} -s}(\mathcal{S}),$$
where the dressed annihilation operators satisfy
$$B_{{\bf p} s}(\mathcal{S})\left|\varphi(\mathcal{S})\right\rangle=0,\quad
D_{{\bf p} s}(\mathcal{S})\left|\varphi(\mathcal{S})\right\rangle=0,$$
and the $D (\mathcal{S})$'s operators can be equivalently seen as creation operators of  negative energy states or hole states belonging to the Dirac sea. All the definitions relative to the vacuum  BCS basis given from Eq. (\ref{QEXP}) to Eq. (\ref{PROJ}) remain valid but with the true vacuum chiral phase $\varphi_p$ replaced by $\varphi(p ;\mathcal{S})$.\\

Formally the in-medium color singlet nucleon state, whose center of mass position is $\bfvec{R}_N$, at  finite local density $\rho (\bfvec{R}_N)$, associated with the field value $\mathcal{S}(\bfvec{R}_N)$, (ignoring for simplicity its explicit spin-color-flavor structure) is generically represented by: 
\begin{equation}
\left|N\right\rangle = V^{3/2}\,\int \frac{d^3 p_1}{(2\pi)^3}\frac{d^3 p_2}{(2\pi)^3}\frac{d^3 p_3}{(2\pi)^3}
\,\Phi\left({\bf p}_1, {\bf p}_2, {\bf p}_3\right)\,B^\dagger_{{\bf p}_1}(\mathcal{S}) B^\dagger_{{\bf p}_2}(\mathcal{S}) B^\dagger_{{\bf p}_3}(\mathcal{S}) \left|\varphi(\mathcal{S})\right\rangle, \label{NUCWF}
\end{equation}
\begin{equation}
\int \frac{d^3 p_1}{(2\pi)^3}\frac{d^3 p_2}{(2\pi)^3}\frac{d^3 p_3}{(2\pi)^3}\,|\left|\Phi\left({\bf p}_1, {\bf p}_2, {\bf p}_3\right)\right|^2 =1.\label{NORM}    
\end{equation}
This state is normalized to 1 ($\left\langle N\right| \left|N\right\rangle =1$) and  $V$ is the volume of a large box. In the following we will explicitly make this construction, the physical underlying physical picture being the in-medium nucleons which look like a Y-shaped string  generated by the non perturbative confining force but with modified constituent quarks at the end. In the following we will most of the time omit for simplicity the explicit dependence on $\mathcal{S}$.\\
\subsection{Bound state equation}
Once the quark fields are expanded on the ($\mathcal{S}$ dependent) BCS basis the confining interaction hamiltonian (Eq.~\ref{HCONF}) can be written with  standard notations ($C=B$ or $D^\dagger$) as: 
\begin{eqnarray}
H_C &=&\frac{1}{2}\,\sum_{k_1 k_2 k_3 k_4}    \langle k_1 k_2|K_C|k_3 k_4\rangle\,
C^\dagger_{k_1}\,C_{k_3}\,C^\dagger_{k_2}\,C_{k_4}	\nonumber\\
&=&\frac{1}{2}\,\sum_{k_1 k_2 k_3 }    \langle k_1 k_2|K_C|k_2 k_3\rangle\,
C^\dagger_{k_1}\,C_{k_3} +\frac{1}{2}\,\sum_{k_1 k_2 k_3 k_4}    \langle k_1 k_2|K_C|k_3 k_4\rangle\,
C^\dagger_{k_1}\,C^\dagger_{k_2}\,C_{k_4}\,C_{k_3}.
\end{eqnarray}
We can expand this interacting hamiltonian using Wick theorem with respect to the BCS-like vacuum $|\varphi(\mathcal{S})\rangle$, up to its quadratic component, namely: 
\begin{equation}
K_C  =K_0\,+\,:K_2:.    
\end{equation}
The vacuum expectation value is ($p$ ($h$) denote positive (negative) energy states):
\begin{eqnarray}
K_0 &=&\langle\varphi(\mathcal{S})|\,H_C\,|\varphi(\mathcal{S})\rangle=\frac{1}{2}\,\sum_{h_1 k_2 }    \langle h_1 k_2|K_C|k_2 h_1\rangle\,
 +\,\frac{1}{2}\,\sum_{h_1 h_2 }   \left( \langle h_1 h_2|K_C|h_1 h_2\rangle\,
-\langle h_1 h_2|K_C|h_2 h_1\rangle\right)\nonumber\\
&=&\frac{1}{2}\,\sum_{h_1 h_2 }\langle h_1 h_2|K_C|h_1 h_2\rangle\,+\,\frac{1}{2}\,\sum_{h_1 p_2 }\langle h_1 p_2|K_C|p_2 h_1\rangle .
\end{eqnarray}
The first term (Hartree term) identically vanishes for a colorless vacuum. Hence:
\begin{eqnarray}
K_0 &=&N_c N_f C_F\,\frac{1}{2} \int\frac{d^3 h_1}{(2\pi)^3}\frac{d^3 p_2}{(2\pi)^3}\int d^3x \,d^3y\, e^{-i({\bf h}_1 - {\bf p}_2)\cdot ({\bf x} - {\bf y})}\,K_C\left({\bf X}, {\bf Y}\right)\nonumber\\
&&Tr_D\left(\gamma_0\,\Lambda_{-}({\bf h}_1)\,\gamma_0\,\Lambda_{+}({\bf p}_2)\right)\equiv 0.
\end{eqnarray}
The $h$ states (hole states) are the negative energy states populating the Dirac sea and the $p$ states are the positive energy states.
The structure of the confining kernel, $K_C\left({\bf X}, {\bf Y}\right)=V_C\left({\bf X}\right)\,+\,V_C\left({\bf Y}\right)$, implies that after spatial integration one automatically gets ${\bf h}_1={\bf p}_2$ and the Dirac trace vanishes, i.e., $Tr_D\left(\gamma_0\,\Lambda_{-}({\bf h}_1)\,\gamma_0\,\Lambda_{+}({\bf h}_1)\right)= 1-c^2_{h_1}-s^2_{h_1}=0$.  Hence the 
confining interaction does not affect the structure of the BCS phase, as already stated in section 2.\\
The quadratic piece, for which  the direct Hartree term does not contribute,   writes 
\begin{eqnarray}
K_2 &=&\left(\frac{1}{2}\,\sum_{k_1 k_2 k_3 }  \langle k_1 k_2|K_C|k_2 k_3\rangle\,-\,\sum_{k_1 k_3 h_2 }    \langle k_1 h_2|K_C|h_2 k_3\rangle\right)\,c^\dagger_{k_1}\,c_{k_3}\nonumber\\
&=&\frac{1}{2}\left(\,\sum_{k_1 p_2 k_3 }  \langle k_1 p_2|K_C|p_2 k_3\rangle\,-\,\sum_{k_1 h_2   k_3 }    \langle k_1 h_2|K_C|h_2 k_3\rangle
\right)\,c^\dagger_{k_1}\,c_{k_3}\nonumber\\
&=& \int d^3x \,d^3y \,\bar{q}(x)\,M({\bf x}, {\bf y})\,{q}(y),
\end{eqnarray}
where M({\bf x}, {\bf y}) is a matrix in Dirac, flavor and color spaces:
\begin{eqnarray}
M({\bf x}, {\bf y})&=&\frac{C_F}{2}\,\gamma_0\,\left[\Lambda_{+}({\bf x}, {\bf y})-\Lambda_{-}({\bf x}, {\bf y})\right]\,\gamma_0\,K_C\left({\bf X}, {\bf Y}\right)\nonumber\\
&=& - C_F\,\,\gamma_0\,\Lambda_{red}({\bf x}, {\bf y})\gamma_0\,K_C\left({\bf X}, {\bf Y}\right),\\
\Lambda_{+}({\bf x}, {\bf y})&=&\int\frac{d^3 p_2}{(2\pi)^3}\, e^{-i{\bf p}_2\cdot ({\bf x} - {\bf y})}\,\Lambda_{+}({\bf p}_2),\nonumber\\
\Lambda_{-}({\bf x}, {\bf y})&=&\int\frac{d^3 h_2}{(2\pi)^3}\, e^{-i {\bf h}_2\cdot ({\bf x} - {\bf y})}\,\Lambda_{-}({\bf h}_2),\nonumber\\
- \Lambda_{red}({\bf x}, {\bf y})&=&-\int\frac{d^3 k}{(2\pi)^3} e^{-i {\bf k}\cdot ({\bf x} - {\bf y})}\Lambda_{red}({\bf k})=\int\frac{d^3 k}{(2\pi)^3} e^{-i {\bf k}\cdot ({\bf x} - {\bf y})}\frac{s_k- c_k\,\bfvec{\gamma}\cdot{\bf{k}}}{2}.
\end{eqnarray} 
It follows that the effect of the inclusion of the confining kernel induced by the presence of the string junction, reduces in a modification of the NJL action by just adding a static mass operator: $\hat{M}(x,y)=M({\bf x}, {\bf y})\,\delta(x_0 - y_0)$. The modified action reads:
\begin{eqnarray}
S &=& S_F\,+\,S_M \\
S_M &=& \int d^4x\,\bigg[-\,\frac{1}{4\,G_1}\,tr_f\left({\cal S}^2\,-\,m\,{\cal S}\,(U+U^\dagger)\right)
		\nonumber\\
  &&+\,\frac{1}{4\,G_2}\,tr_f\left(({V}^\mu 
  +{\cal V}^\mu_c )^2\,+\,({A}^\mu +{\cal A}^\mu_c )^2\right)\bigg]\\
S_F &=& \bar{q}\, D \,q = \bar{q}\,	\left( D_{NJL}\,-\,\hat{M}\right)\,q\\
&\equiv& \int d^4x\,\bar{q}(x)\left[i\,\gamma^{\mu}\partial_\mu-{\cal S}-\gamma^{\mu}\left( V_\mu+\gamma_5\, A_\mu\right)\right](x)q(x)\nonumber\\
&&-\,
\int d^4x d^4y\,\bar{q}(x)M({\bf x}, {\bf y})\,\delta(x_0-y_0) q(x).
\end{eqnarray}
To leading order in the pion field, the fermionic piece of the NJL action,  involving the dominant p-wave coupling of the pion, reduces to: 
	\begin{eqnarray}
		\bar{q}\,	 D_{NJL}\,q &=&\int d^4x \,\bar{q}(x)\left[	i\gamma_{\mu}\left(\partial^{\mu}+ i \mathcal{V}_c^{\mu}\right) - \gamma_{\mu}\gamma^{5} \mathcal{A}_c^{\mu} - \mathcal{S}\right]q(x)\nonumber\\
		&\simeq&\int d^4x \,\bar{q}(x)\left[	i\gamma_{\mu}\partial^{\mu} - \mathcal{S} - \gamma_{\mu}\gamma^{5} \mathcal{A}_c^{\mu} \right]q(x)\nonumber\\
		&\simeq&\int d^4x\, \bar{q}(x)\,\left[i\gamma_{\mu}\partial^{\mu}  - \mathcal{S} -\frac{1}{F_\pi}\gamma^j\partial_j\vec{\Phi}\cdot\vec{\tau} \right]q(x).
\end{eqnarray}
For fixed  given  meson fields considered as external fields, and in particular for $\mathcal{S}$, adjusted to their  realization in the "shifted vacuum"  at a nuclear density $\rho$, the partition function, generating functional of the in-medium confined quark Green's functions, is :
\begin{eqnarray}
Z(\eta,\bar{\eta};\mathcal{S})&=& \langle\varphi(\mathcal{S})|e^{-i \left[H -d^3x (\bar{q}\eta+\bar{\eta}q)(x)\right]\,T}\,|\varphi(\mathcal{S})\rangle\nonumber\\
&=& \int d\bar{q}(x) dq(x)\, e^{i\bar{q}\, D \,q\,+\,id^4x \,(\bar{q}\eta+\bar{\eta}q)(x)\,+\,i S_M}\nonumber\\
&=& exp\left(-i Tr\,\ln D \,-i\, \bar{\eta}\, D^{-1}\,\eta \,+\,iS_M\right).
\end{eqnarray}
Let us introduce the quark propagator (i.e., the time-ordered one-particle Green's function): 
\begin{eqnarray}
S_{\alpha,\alpha'}^{af,a'f'}(x,x')&=&  \langle\varphi(\mathcal{S})| -i\, \mathcal{T}\left(q_{\alpha}^{af}(x), \bar{q}_{\alpha'}^{a'f'}(x')\right)|\varphi(\mathcal{S})\rangle\nonumber\\
&=&\langle\varphi(\mathcal{S})|-i\, \mathcal{T}\left(q_{\alpha}^{af}({\bf{x}},t), \bar{q}_{\alpha'}^{a'f'}({\bf{x}}',t')\right)|\varphi(\mathcal{S})\rangle.  
\end{eqnarray}
It can be obtained from the partition function according to:
\begin{equation}
S_{\alpha,\alpha'}^{af,a'f'}(x,x')=i\frac{\delta}{\delta \eta_{\alpha}^{af}(x)} \frac{\delta}{\delta \bar{\eta}_{\alpha'}^{a'f'}(x')} Z(\eta,\bar{\eta};\mathcal{S})= \left[D^{-1}\right]_{\alpha,\alpha'}^{af,a'f'}(x,x').
\end{equation}
Hence one obtains a Dyson-Schwinger equation: 
\begin{eqnarray}
\int d^4z\, D_{\alpha,\beta}^{af,bg}(x,z)\,S_{\beta,\alpha'}^{bg,a'f'}(z,x')=\delta_{aa'}\,\delta_{ff'}\,\delta^{(4)}(x-x').
\end{eqnarray}
Still keeping the coupling to the pion field, its explicit form for a static mass operator reads:
\begin{eqnarray}
&&\left[\left(i\gamma_0\frac{\partial}{\partial t}\,+\, \bfvec{\gamma}\cdot\bfvec{\nabla}_{x} \,-\,\mathcal{S}\right)\,-\frac{1}{F^2_\pi}\gamma^j\partial_j\vec{\Phi}(\bfvec{x}, t)\cdot{\vec\tau} \right]^{fg}\,S_{\alpha,\alpha'}^{ag,a'f'}(\bfvec{x}, t;  \bfvec{x}', t')\nonumber\\
&& -\,\int d{\bf{z}}\,\,M(\bfvec{x}, \bfvec{z})\,\,S_{\alpha,\alpha'}^{af,a'f'}(\bfvec{z}, t;  \bfvec{x}', t')=\delta_{aa'}\,\delta_{ff'}\,\delta^{(3)}(\bfvec{x}\,-\,\bfvec{x}')\,\delta(t-t').\label{SDE}
\end{eqnarray}
\\
We now derive a  Dirac-like equation for the confined quark wave function. For that purpose we ignore the pion coupling which will be reintroduced later in perturbation theory in a way similar to the cloudy bag model (its contribution to the nucleon mass, namely the pionic self-energy, has  already been given in Eq. (\ref{SELFENERGY}) in subsection 4.5). The quark propagator can be expressed in term of the Dirac eigenfunctions, $$\left(\psi_n\right)_{\alpha}^{af}({\bf{x}})=\langle\varphi(\mathcal{S})|\,q_{\alpha}^{af}({\bf{x}})\,|\,n\,\rangle,$$ 
according to :
\begin{equation}
 S_{\alpha,\alpha'}^{af,a'f'}({\bf{x}}, t;  {\bf{x}}', t') =\int \frac{d\omega}{2\pi}\,e^{-i\omega(t-t')}\,\sum_n\left[\frac{\left(\psi_n\right)_{\alpha}^{af}({\bf{x}})\left(\bar\psi_n\right)_{\alpha'}^{a'f'}({\bf{x}}')}{\omega\,-\,E_n\,+\,i\eta} +\frac{\left(\bar\psi_n\right)_{\alpha'}^{a'f'}({\bf{x}}')\left(\psi_n\right)_{\alpha}^{af}({\bf{x}})}{\omega\,+\,E_n\,-\,i\eta}\right].
\end{equation}
Taking the Fourier transform in energy space of the Dyson-Schwinger equation (\ref{SDE}), multiplying both sides by $\omega -E_n$ and taking the limit $\omega \to E_n$, one obtains an eigenvalue equation for the confined bound state described by the Dirac wave function, $\left(\psi_n\right)({\bf{x}})$, which is independent of color and flavor if we ignore the up-down current quark mass difference:
\begin{equation}
\left[-i\bfvec{\alpha}\cdot\vec\nabla_{\bf{x}}+ \beta\, \mathcal{S}\right] \left(\psi_n\right)({\bf{x}}) +  \int d{\bf{z}}\,\beta\,M({\bf x}, {\bf z})\,\left(\psi_n\right)({\bf{z}})  =E_n \left(\psi_n\right)({\bf{x}}).
\end{equation}
In case of full chiral symmetry breaking, $\varphi_k=\pi/2$, i.e., for very large constituent quark mass, the mass operator reduces to $M({\bf x}, {\bf z})=C_F\,V_C({\bf x})\,\delta^{(3)}(\bf{x}\,-\,\bf{z})$. Hence we arrive at the conclusion that this local confining potential is predominantly of scalar nature, although the starting color interaction is of vector nature. Hence dynamical chiral symmetry breaking generates naturally scalar confinement, in agreement with phenomenology at variance with approaches including by hand an explicit chiral symmetry breaking to artificially generate scalar confinement. This point is abundantly discussed in the literature in the context of heavy-light $Q\bar{q}$ systems, see e.g., Ref. \cite{BBRV98,KNR2017}. This scalar confinement has also the expected long range linear behaviour $V_S(R)=\beta C_F\,V_C(R)\simeq \beta\sigma R$ for $R>2T_g/\sqrt(\pi)\simeq 0.4\,fm$. Hence this scalar potential varies very slowly inside the quark core (inside the "bag") and linearly increases beyond. This picture resembles the MIT bag for which confinement is ensured by a constant tiny  scalar potential inside the bag and an infinite scalar potential outside of the bag. \\
Using the expansion of the quark field on the BCS basis, the Dirac wave function takes the form: 
\begin{equation}
\left(\psi_n\right)({\bf{x}})=\int\frac{d^3 p}{(2\pi)^3} \,  e^{-i {\bf p}\cdot {\bf x}}\, (\tilde\psi_n)({\bf{p}}).
\end{equation}
The Dirac wave function in momentum space has a specific quasi plane wave (QPW) form,
\begin{equation}
 (\tilde\psi_n)({\bf{p}})=   =\sqrt{\frac{1}{2}}\left(
\begin{array}{c}
	\sqrt{1+s_p}\,\,\Phi_n({\bf{p}})\\
	\sqrt{1-s_p}\,\,\bfvec{\sigma}\cdot \hat{\bf p}\,\,\Phi_n({\bf{p}})
\end{array}\right)
=\sqrt{\frac{E_P + \mathcal{S}}{2 E_P}}\left(
\begin{array}{c}
	\Phi_n({\bf{p}})\\
	\frac{\bfvec{\sigma}\cdot {\bf p}}{E_P + \mathcal{S}}\,\,\Phi_n({\bf{p}})
\end{array}\right),
\end{equation}
where the two-component Pauli spinor wave function is:
\begin{equation}
  \Phi_n({\bf{p}})=\sqrt{V}\,\sum_s \langle\varphi(\mathcal{S})|\,B_{{\bf{p}}s}(\mathcal{S})\,|\,n\,\rangle\,\,\chi_s.
\end{equation}
We are looking for normalized single quark states with well defined parity and total angular momentum:
\begin{eqnarray}
|n;ljm\rangle &=&\frac{1}{\sqrt{V}}\sum_{{\bf{p}}}\sqrt{4}\pi\,R_{lj}(p)\,\sum_{\mu s} \,\langle l\mu, \frac{1}{2}s|jm\rangle \,Y_{l\mu}\,(\hat{\bf p})\,B^\dagger_{{\bf{p}}s}(\mathcal{S})\,|\varphi(\mathcal{S})\rangle .
\end{eqnarray}
Hence the spinor wave function writes:
\begin{eqnarray}
\Phi_{n;ljm}({\bf{p}})&=&\sqrt{4}\pi\,R_{lj}(p)\,\Phi_{ljm}(\hat{\bf{p}})\nonumber\\
\Phi_{ljm}(\hat{\bf{p}})&=&\sum_{\mu s} Y_{l\mu}(\hat{\bf p})\,\chi_s\,\langle l\mu, \frac{1}{2}s|jm\rangle\quad \hbox{with} \quad {\bfvec{\sigma}}\cdot\hat{\bf{p}}\,\Phi_{ljm}=-\Phi_{l'jm},\,\, l'=2j-l.
\end{eqnarray}
The normalization of the quark state implies: 
\begin{equation}
 \int\frac{d^3 p}{(2\pi)^3} \,|\Phi_{n;ljm}|^2  ({\bf{p}})=\int\frac{d^3 p }{(2\pi)^3} \,  R^2_{lj}(p)=1.
\end{equation}
Passing in momentum space we obtain the  Dirac eigenvalue equation,
\begin{equation}
\left[\bfvec{\alpha}\cdot{\bf p}\,+ \,\beta\, \mathcal{S}\right] (\tilde\psi_{n;ljm})({\bf{p}}) \,+\,\int \frac{d^3q}{(2\pi)^3}\beta,\tilde{M}({\bf p},{\bf q}) \,(\tilde\psi_{n;ljm})({\bf{q}})=E_{lj}\,(\tilde\psi_{n;ljm})({\bf{p}}),
\end{equation}
with: 
\begin{eqnarray}
\tilde{M}({\bf p},{\bf q})&=&\int d^3x\,d^3y\,e^{-i({\bf p}\cdot {\bf x} - {\bf q}\cdot{\bf y})}\,M({\bf x}, {\bf y}).
\end{eqnarray}
At this level it is convenient to introduce an effective one-body confining potential $W(R)$ such that: 
\begin{eqnarray}
 C_F K_C\left({\bf X}, {\bf Y}\right)&=&C_F V_C\left({\bf X}\right)\,+\,C_F V_C\left({\bf Y}\right)=W_C\left({\bf X}\right)+ W_C\left({\bf Y}\right)\nonumber\\
W_C\left({\bf R}\right)&=& W_S\left({\bf R}\right)\,-\,\frac{2\,\sigma\,T_g}{\sqrt{\pi}}.
\end{eqnarray}
The effective string potential is
\begin{equation}
W_S\left({\bf R}\right)=\frac{\sigma}{2\,\sqrt{\pi}\,T_g}\,{\bf R}^2
\,I\left({\bf R},{\bf R}\,;\,0\right)=\,\frac{\sigma}{2\,\sqrt{\pi}\,T_g}\,{\bf R}^2
\,\int_0^1 dv\,\int_0^1 dw\,e^{-\left(\left(v-w\right)^2\frac{{\bf R}^2}{4 T^2_g}\right)},
\end{equation}
with asymptotic behavior:
\begin{equation}
R<<T_g :  \quad 
W_S\left({\bf\ R}\right)=\frac{\sigma}{2\,\sqrt{\pi}\,T_g}\, R^2, \qquad
R>>T_g : \quad W_S\left({\bf R}\right)= \,\sigma\,\left(R\,-\,\frac{2}{\sqrt{\pi}}\,T_g\right).
\end{equation}
Actually this potential deviates from the above $V_C({\bf R})$ by a factor $1/C_F$ with the constant shift $2\sigma T_g/\sqrt{\pi}$ removed.
Introducing the projector on the QPW Dirac spinors, 
$$U_{\bf p}=\beta\, \sin\varphi_p\,+\,\,\bfvec{\alpha}\cdot\hat{\bf p}\,\cos\varphi_p$$
the mass operator takes the form:
\begin{eqnarray}
\beta\tilde{M}({\bf p},{\bf q})&=&\frac{1}{2}\,\int d^3R\,e^{-i({\bf p} - {\bf q})\cdot{\bf R}}\,W_C({\bf R})\,\left(U_{\bf p}\,+U_{\bf q}\right)\equiv\frac{1}{2}\,\tilde{W}_C({\bf p} - {\bf q})\,\left(U_{\bf p}\,+U_{\bf q}\right)\nonumber\\
&=&\frac{1}{2}\,\tilde{W}_S({\bf p} - {\bf q})\,\left(U_{\bf p}\,+U_{\bf q}\right)\,-\,(2\pi)^3\delta^{(3)}\left({\bf p} - {\bf q}\right)\,\frac{2\sigma}{\,\sqrt{\pi}\,T_g}\,U_p.
\end{eqnarray}
Consequently the Dirac-like bound state equation writes
\begin{eqnarray}
&&E_p\,U_p\,(\tilde\psi_{n;ljm})({\bf{p}})\,+\,\frac{1}{2}\,\int\frac{d^3 q}{(2\pi)^3}\,\tilde{W}_C({\bf p} - {\bf q})\,\left(U_{\bf p}\,+U_{\bf q}\right)(\tilde\psi_{n;ljm})({\bf{q}})\\
&&=E_{jl}\,(\tilde\psi_{n;ljm})({\bf{p}})\\
&&E_p\,U_p\,(\tilde\psi_{n;ljm})({\bf{p}})\,+\,\frac{1}{2}\,\int\frac{d^3 q}{(2\pi)^3}\,\tilde{W}_S({\bf p} - {\bf q})\,\left(U_{\bf p}\,+U_{\bf q}\right)(\tilde\psi_{n;ljm})({\bf{q}})\nonumber\\
&& =\left(E_{jl}\,+\,\frac{2\sigma}{\sqrt{\pi}\,T_g}\,U_p\right)\,(\tilde\psi_{n;ljm})({\bf{p}}),
\end{eqnarray}
which is just the equation (111) of Ref. \cite{KNR2017}. It is simple matter to show that this equation is equivalent to a Schr\"{o}dinger-like bound state equation for the two-component spinor $\Phi_{n;ljm}({\bf{p}})$,
\begin{eqnarray}
&&E_p\,\Phi_{n;ljm}({\bf{p}})\,+\,\frac{1}{2}\,\int\frac{d^3 q}{(2\pi)^3}\,\tilde{W}_C({\bf p} - {\bf q})\,\bigg(\sqrt{1+s_p}\sqrt{1+s_q}\\\nonumber
&&+\sqrt{1-s_p}\sqrt{1-s_q}\,\bfvec{\sigma}\cdot {\bf p}\,\bfvec{\sigma}\cdot {\bf q}\bigg)\,\Phi_{n;ljm}({\bf{p}})=E_{jl}\,\Phi_{n;ljm}({\bf{p}})\\
&&E_p\,\Phi_{n;ljm}({\bf{p}})\,+\,\frac{1}{2}\,\int\frac{d^3 q}{(2\pi)^3}\,\tilde{W}_S({\bf p} - {\bf q})\,\bigg(\sqrt{1+s_p}\sqrt{1+s_q}\nonumber\\
&&+\sqrt{1-s_p}\sqrt{1-s_q}\,\bfvec{\sigma}\cdot {\bf p}\,\bfvec{\sigma}\cdot {\bf q}\bigg)\,\Phi_{n;ljm}({\bf{p}})=\left(E_{jl}\,+\,\frac{\sigma}{2\,\sqrt{\pi}\,T_g}\right)\,\Phi_{n;ljm}({\bf{p}}),
\end{eqnarray}
where $\tilde{W}_C({\bf p})$ ($\tilde{W}_S({\bf p})$) is the Fourier transform of the  effective one-body confining potential $W_C({\bf R})$ ($W_S({\bf R})$). To establish the above results we have used a specific property of the QPW wave function, namely:
$$U_{\bf p}(\tilde\psi_{n;ljm})({\bf{p}})=\left(\beta\, \sin\varphi_p\,+\,\,\bfvec{\alpha}\cdot\hat{\bf p}\,\cos\varphi_p\right)(\tilde\psi_{n;ljm})({\bf{p}})=(\tilde\psi_{n;ljm})({\bf{p}}).$$
It follows that the bound state energy can written as well as:
\begin{equation}
E_{jl}=\int d^3 x\, \left(\psi_n\right)(\bfvec{x})\left[-i\bfvec{\alpha}\cdot\vec\nabla_{\bfvec{x}}\,+\, \beta M \, +  \,\,W_C(\bfvec{x})\right]\,\left(\psi_n\right)(\bfvec{x}).    
\end{equation}
Hence the problem of the determination of the bound state  is equivalent to solve a Dirac equation for a quark moving in an effective central potential $W_C(x)$, but  with solutions limited to the  subset of QPW wave functions. The final form of the bound state energy follows,
\begin{eqnarray}
E_{jl}&=&\int d^3 x\, \left(\psi_n\right)(\bfvec{x})\left[-i\bfvec{\alpha}\cdot\vec\nabla_{\bfvec{x}}\,+\, \beta M \, +  \,\,W_S(\bfvec{x})\right]\,\left(\psi_n\right)(\bfvec{x})-\,\frac{2\sigma T_g}{\sqrt{\pi}}\nonumber\\
&\equiv & E_{jl, kin}+E_{jl, pot}, \nonumber\\
E_{jl, kin}&=&\int \frac{d^3p}{(2\pi)^3}\,E_p\,R^2_{jl}(p), \nonumber\\
E_{jl, pot}&=&\frac{1}{2}\,\int d^3r\, W_S(r)\,\left(|G_{1l}(r)|^2 + |G_{2l'}(r)|^2\right)\,-\,\frac{2\sigma T_g}{\sqrt{\pi}},
\end{eqnarray}
with:
\begin{equation}
  G_{1l}(r)=\int \frac{d^3q}{(2\pi)^3}\,\sqrt{1+s_q}\,R_{jl}(q)\,j_l(qr),\quad
G_{2l'}(r)=\int \frac{d^3 q}{(2\pi)^3}\,\sqrt{1-s_q}\,R_{jl'}(q)\,j_{l'}(qr).  
\end{equation}
We look for a Gaussian solution for the lowest orbital with $j=1/2,\, l=0,\,l'=1 $, 
\begin{equation}
R_0(p)= (2\pi)^{\frac{3}{2}}\,\left(\frac{b^2}{\pi}\right)^{\frac{3}{4}}\,e^{-b^2\,p^2/2},  
\end{equation}
$b$ being a variational parameter. One can search a solution for any value $M (s)$ of the in-medium constituent quark mass hence producing ultimately a density dependent nucleon mass as a function of the value of the "nuclear physics sigma meson" field $s$. \\
If we first consider the case where the nucleon wave function is just the product of the three quark orbitals properly projected on to the color singlet state with $I=J=1/2$, namely,
\begin{equation}
\Phi\left({\bf p}_1, {\bf p}_2, {\bf p}_3\right)= \Phi\left({\bf p}_1\right)\,  
\Phi\left({\bf p}_2\right)\,  
\Phi\left({\bf p}_3\right)=R_0(p_1) R_0(p_2) R_0(p_3),
\end{equation}
the mass of the nucleon ignoring pion cloud and gluon exchange correction is simply:
\begin{equation}
M_N^{(core, \,no\, CM\, correction)}(\mathcal{S})=3\,\left(E_{0, kin}(\mathcal{S}) + E_{0, pot}(\mathcal{S})\right)
\equiv 3\,E_0(\mathcal{S}).
\end{equation}
An approximate way to eliminate the spurious center of mass energy is to take the nucleon mass as being given by:
\begin{equation}
M_N^{(core)}(\mathcal{S})=\sqrt{9E_0^2(\mathcal{S}) -\langle P^2\rangle(\mathcal{S})} = 3\,\sqrt{E_0^2(\mathcal{S}) -\frac{1}{2 b^2(\mathcal{S})}}.
\end{equation}
If we limit ourselves to the leading order response parameters, all what we need is the mass and its first and second derivatives taken at the vacuum value $\mathcal{S}=M_0$:
\begin{equation}
g_S=\frac{M_0}{F_\pi}\left(\frac{\partial M_N^{(core)}(\mathcal{S})}{\partial\mathcal{S} }\right)_{\mathcal{S}=M_0},\qquad
C=\frac{M_0^2}{2 M_N}\left(\frac{\partial^2 M_N^{(core)}(\mathcal{S})}{\partial\mathcal{S}^2 }\right)_{\mathcal{S}=M_0}.
\end{equation}
\subsection{Results for the NJLSET1 set of parameters}
With the QCD inputs corresponding to the NJLSET1 (i.e., ignoring pion-axial mixing, $G_2=0$), $\sigma=0.18\,\mathrm{GeV}^2$ and $\mathcal{G}_2=0.025\,\mathrm{GeV}^4$ yielding   $T_g= 0.286\,\mathrm{fm}$, $\eta=\sqrt{\sigma}T_g=0.615$ and $M_0=356.7\,\mathrm{MeV}$, the value of $b$  minimizing the orbital energy is $b=0.978\,\sqrt{\sigma}$.  The resulting contribution to the quark  orbital energy  is $E_0=E_{0, kin} + E_{0, pot}=1.458\,\sqrt{\sigma}\,+\,0.636\,\sqrt{\sigma} -\,\,0.694\,\sqrt{\sigma}$, where the last contribution comes from the constant shift. After center of mass correction the quark core contribution to the nucleon mass is:
\begin{equation}
M_N^{(core)}=\sqrt{9E_0^2 -\langle P^2\rangle} = 3\,\sqrt{E_0^2 -\frac{1}{2 b^2}}=3.62\,\sqrt{\sigma}.
\end{equation}
For the response parameters we numerically obtain:
\begin{eqnarray}
g_S^{(core, NJLSET1)}&=&\frac{M_0}{F_\pi}\left(\frac{\partial M_N^{(core)}(\mathcal{S})}{\partial\mathcal{S} }\right)_{\mathcal{S}=M_0}=6.48\\
C^{(core, NJLSET1)}&=&\frac{M_0^2}{2 M_N}\left(\frac{\partial^2 M_N^{(core)}(\mathcal{S})}{\partial\mathcal{S}^2 }\right)_{\mathcal{S}=M_0}=0.30.
\end{eqnarray}
The value of the $C$ parameter looks {\it a priori} small but since $g_S$ is significantly smaller than $M_N/F_\pi$, its effective value, $(M_N/g_S F_\pi)\, C$, occuring in Eq.~(\ref{THREEBODN}) and  Eq.~(\ref{LATTIX}) is larger, e.g, $(M_N/g_S F_\pi)\, C =0.50$. If we include the value of $C_\chi=0.488$ also issued from the microscopic FCM-NJL approach, one obtains for the parameter $\Tilde C_3$ entering the three-body force (Eq.~\ref{THREEBODN}) and $\Tilde{C}_L$ (Eq.~\ref{LATTIX}) entering the chiral susceptibility $a_4$, one has:
\begin{equation}
 \Tilde{C}_3=\frac{M_N}{g_S F_\pi}\,C +\frac{1}{2}\,C_\chi=0.73,\qquad
 \Tilde{C}_L=\frac{M_N}{g_S F_\pi}\,C +\frac{3}{2}\,C_\chi=1.21.
\end{equation}
It turns out that the $s$ field whose value lies in the range $[0 \div-F_\pi]$ does not exactly coincide with the canonical scalar field, $s_c$, of the bosonised NJL lagrangian. According to Eq. (\ref{CANON}), they are related by 
$s=z_S\,s_c$ where the dimensionless rescaling factor $z_S$ is expressible in terms of NJL loop  integrals. With the above NJL parameters its numerical value is $z_S=1.284$. Hence the values of the scalar coupling constant and of the scalar mass associated to the canonical scalar field are: 
\begin{equation}
 g_{\sigma c} =z_S\,g_S=8.37, \qquad  m_{\sigma c} =z_S\, M_\sigma=919\,\mathrm{MeV}.
\end{equation}
This is the reason for which we quote those values of $g_{\sigma c}$ and $m_{\sigma c}$ in the parameters entering the $NN$ potential in Eq. (43) of Ref. \cite{Universe}. Also notice that the ratio, $g_S/M_\sigma=g_{\sigma c} /m_{\sigma c} $, is not affected when passing  from the field $s$ to the canonical scalar field $s_c$ and this rescaling has no effect at the Hartree level but may have a small effect for the Fock terms.
It is also possible to calculate the core rms and the vertex form factors in the various (scalar, axial and vector) Yukawa channels. We give here the cutoff values for the equivalent monopole form factors regularizing the corresponding Yukawa couplings to the nucleon:
\begin{equation}
 \Lambda_\sigma\simeq 0.85\,GeV,\qquad   \Lambda_\pi\simeq 1.\,GeV,\qquad  \Lambda_v\simeq 1.1\,GeV.
\end{equation}
This simple estimate demonstrates that they are close to the $\Lambda =1\,GeV$  in the NN potential (Eq. (43) of \cite{Universe}).\\
Finally in  view of the next section dedicated to the pion cloud contribution to the nucleon mass, let us mention that the nucleon axial coupling constant is $g_A=1.24$.
\subsection{Results for the NJLSET2 set of parameters}
With the QCD inputs corresponding to the NJLSET2 ( $G_2=0.75\,G_1=7.5\,\mathrm{GeV}^{-2}$), $\sigma=0.18\,\mathrm{GeV}^2$ and $\mathcal{G}_2=0.039\,\mathrm{GeV}^4$ yielding   $T_g= 0.229\,\mathrm{fm}$, $\eta=\sqrt{\sigma}T_g=0.492$ and $M_0=358\,\mathrm{MeV}$, the value of $b$  minimizing the orbital energy is $b=0.95\,\sqrt{\sigma}$.  The resulting contribution to the quark  orbital energy  is $E_0=E_{0, kin} + E_{0, pot}=1.488\,\sqrt{\sigma}\,+\,0.708\,\sqrt{\sigma} -\,\,0.554\,\sqrt{\sigma}$. This value is significantly larger than in the first set parameters. This is due to a smaller value of the depth of the potential, $-2\sigma T_g/\sqrt{\pi}$. After center of mass correction the quark core contribution to the nucleon mass is unrealistically large:
\begin{equation}
M_N^{(core)}=\sqrt{9E_0^2 -\langle P^2\rangle} = 3\,\sqrt{E_0^2 -\frac{1}{2 b^2}}=4.44\,\sqrt{\sigma}.
\end{equation}
But for what concern the response parameters, the results,
\begin{eqnarray}
g_S^{(core, NJLSET2)}&=&\frac{M_0}{F_\pi}\left(\frac{\partial M_N^{(core)}(\mathcal{S})}{\partial\mathcal{S} }\right)_{\mathcal{S}=M_0}=6.54\\
C^{(core, NJLSET2)}&=&\frac{M_0^2}{2 M_N}\left(\frac{\partial^2 M_N^{(core)}(\mathcal{S})}{\partial\mathcal{S}^2 }\right)_{\mathcal{S}=M_0}=0.299,
\end{eqnarray} 
 are almost indistinguishable from the one of the first set of parameters.
 Let us also mention that the axial coupling constant, $g_A=1.23$ is also very close to the one of the NJLset1 set of parameters.

 Hence we can conclude that, except for the value of the potential pocket, the two sets of parameters (the second one including the vector NJL coupling, $G_2)$) give essentially indistinguishable predictions for what concerns the in-medium evolution of the nucleon mass and especially the response parameters $g_S$ and $\kappa_{NS}$. 
 For that reason in the following we will limit our discussion to the first set of parameters, NJLset1. 
\section{Pion cloud contribution to the in-medium nucleon mass}
The explicit calculation of the pionic self-energy (Eq. \ref{SELFENERGY}) generates  a contribution to the free nucleon mass:
$$
M_N^{(pion\,cloud)}=\Sigma^{(\pi)}({\mathcal{S}=M_0 ;m})=-1.152\,\sqrt{\sigma}.
$$
If we add a gluon-exchange correction calculated as in \cite{Jena97},
$$
M_N^{(gluon\,exch)}=-1.37\,\alpha_S\sqrt{\sigma},
$$
and taking $\sigma=0.18\,GeV^2$, one can adjust the effective strong coupling constant as $\alpha_S=0.625$ (close to $\alpha_S=0.576$ used in \cite{Jena97}) to obtain:
$$
M_N=M_N^{(core)}+M_N^{(pion\,cloud)}+M_N^{(gluon\,exch)}= 1534\,-\,488\,-\,109=938\,\mathrm{MeV}.
$$
This is indeed an encouraging result even if the magnitude of the core contribution and of the pion cloud contribution seem separately too large. We expect that a more thorough account of the CM correction and/or the nucleon stability will be able to improve this aspect; this will be discussed below. The most serious problem shows up when we consider the pion cloud contribution to the response parameters. For $g_S$, we find: 
$$
g_S^{(pion\,cloud)}=\frac{M_0}{F_\pi}\left(\frac{\partial M_N^{(pion\,cloud)}(\mathcal{S})}{\partial\mathcal{S} }\right)_{\mathcal{S}=M_0}=-0.57.
$$
At first glance, one may think that this rather moderate value reflects that the pion properties are protected by chiral symmetry, as it is often assumed. However the contribution to the $C$ parameter,
$$
C^{(pion\,cloud)} = \frac{M_0^2}{2 M_N}\left(\frac{\partial^2 M_N^{(pion\,cloud)}(\mathcal{S})}{\partial\mathcal{S}^2 }\right)_{\mathcal{S}=M_0}=-0.30,
$$
is large and negative, just compensating the core contribution! 
Even though at normal density, for which $s/F_\pi\sim-0.2$, the generated  extra attraction, $\delta (E/A)^{(C,\,pion)}\sim -13\,\mathrm{MeV}$, is partially compensated by the extra repulsion, $\delta (E/A)^{(g_S,\,pion)}\sim 10.5\,\mathrm{MeV}$, the original saturation mechanism is certainly severely affected. \\

To identify the physical origin of this effect, let assume first  that, for some reason (such as "pion properties are protected by chiral symmetry"), the only medium effects affecting the nucleon mass are those
generating a readjustment of the quark wave function in presence of the nuclear scalar field, exactly as in the QMC model. This is equivalent to freeze in the expression of the pionic self-energy the in-medium pion mass, $M_\pi(\mathcal{S})$, and the in-medium pion decay constant, $F_\pi(\mathcal{S})$, to their vacuum values. Hence the only effect affecting the pionic-self energy comes from the evolution with $\mathcal{S}$ of the axial coupling constant and of the axial form factor (Eq. \ref{GAC}). In such a case, the contribution of the pion cloud to the $C$ parameter becomes very small: $C^{(pion\,cloud)}=-0.02$ and the associated  repulsive three-body force at the origin of the saturation mechanism (Eq.~\ref{THREEBODN}) is preserved. However the contribution to the scalar coupling constant is rather large and negative, $g_S^{(pion\,cloud)}=-3.55$,  reducing  by more than $50\%$ the quark core value. This  of course will  generate the problem of not having enough attraction to bind  nuclear matter and difficulties with the magnitude of the spin-orbit potential (see a recent
discussion for this specific point in Ref. \cite{spinorbit}). \\
 
There is another more founded source of improvement which is associated with the necessity to solve the (very old) problem of the lack of nucleon stability in the "chiral bag" models \cite{Brown}. This is indeed the case  in our model since the pion self-energy  behaves as 
$$\Sigma^{(\pi)}({\mathcal{S};m})\sim -\left(\frac{g_A}{2 F_\pi}\right)^2\,\frac{1}{b^3},$$
where $b$ represents the size of the quark core. Consequently, as the size of the quark core decreases, the negative pion pressure may become very large surpassing the Fermi pressure, causing a possible collapse of the "quark core bag". The origin of the problem can be identified if one realizes that the elementary pion field couples derivatively  to the quarks mainly at the surface (i.e., $r_c\sim b\sim 1/\sqrt{\sigma}\sim 0.5\,fm$ ) of the quark core, which is not larger that the physical pion size extracted from the the pion electromagnetic form factor, $\sqrt{\langle r^2_\pi\rangle} =0.65\,fm$. In the NJL model this rms radius (keeping only the dominant three-quark loop contribution, see section VI-4 of Ref. \cite{KLE}) is such that: 
$$
\langle r^2_\pi\rangle_{NJL}=\frac{3}{4\pi^2 F^2_\pi}=0.34\,fm^2\simeq(0.6\, fm)^2.
$$
Hence, the finite pion size cannot be ignored in the calculation of the pionic self-energy. To incorporate its effect, we replace the local coupling of the elementary pion field by a non local coupling as in Ref. \cite{Pirner},
\begin{equation}
 \mathcal{L}_{\pi qq}= \bar{q}({\bf x})\,\gamma^j\gamma^5\, q({\bf x})\,\partial_j\Phi({\bf x})
\quad \to\quad \int d^3\eta\,\bar{q}({\bf x}+ \bfvec{\eta})\,\gamma^j\gamma^5\, q({\bf x} -\bfvec{\eta})\,P(\bfvec{\eta})\,\partial_j\Phi({\bf x}),
\end{equation}
where $P(\bfvec{\eta})$ gives the probability to find a quark and antiquark at relative distance $2\eta$ in the extended pion. We take for $P(\bfvec{\eta})$ a  gaussian form:
$$P(\bfvec{\eta})=\left(\frac{1}{\pi\,\rho^2_\pi(\mathcal{S})}\right)^{3/2}\,exp\left(-\frac{\eta^2}{\rho^2_\pi(\mathcal{S})}\right).$$
We expect the size parameter $\rho_\pi$ to be directly related to the electromagnetic size of the pion. Since this pion radius comes from a constituent quark loop, we expect that it depends on the in-medium constituent quark mass, $M=\mathcal{S}$ and consequently on the in-medium pion decay constant parameter $F_\pi(\mathcal{S})$. For orientation we  take:
\begin{equation}
\rho^2_\pi(\mathcal{S})= \frac{1}{4\pi^2 F^2_\pi(\mathcal{S})}.
\end{equation}
It follows that the size of the composite pion progressively increases with increasing density. In other words the pion progressively dilutes as a consequence of partial chiral symmetry restoration.  The net effect of the finite pion size is to replace the axial form factor (Eq. \ref{GAC}) appearing in the expression of the self-energy (\ref{SELFENERGY}) by a $\pi NN$ form factor, according to:
$$
v({\bf q}; \mathcal{S})\qquad\to\qquad exp\left(-\frac{q^2\,\rho^2_\pi(\mathcal{S})}{4}\right)\,v({\bf q}; \mathcal{S}).
$$
Very qualitatively, the inclusion of the scalar field dependent pion size, modifies the $1/b^3(\mathcal{S})$ behavior of the pionic self-energy to $1/[(b^2(\mathcal{S})+\rho^2_\pi(\mathcal{S})]^{3/2}$.
Once the finite pion size effect is taken into account, the contribution of the pion cloud to the response parameters becomes:
$$
g_S^{(finite\,size\,pion\,cloud)}=\frac{M_0}{F_\pi}\left(\frac{\partial M_N^{(pion\,cloud)}(\mathcal{S})}{\partial\mathcal{S} }\right)_{\mathcal{S}=M_0}=-0.94,
$$
$$
C^{(finite\,size\,pion\,cloud)} = \frac{M_0^2}{2 M_N}\left(\frac{\partial^2 M_N^{(pion\,cloud)}(\mathcal{S})}{\partial\mathcal{S}^2 }\right)_{\mathcal{S}=M_0}=-0.02.
$$
The main result is that the contribution to the $C$ parameter is now very small. Hence the inclusion of the pion cloud do not significantly affects the original saturation mechanism linked to the polarization of the quark core in presence of the nuclear scalar field. We also show on figure 2 the result of the evolution of the nucleon mass namely, $\Delta M_N(\mathcal{S})=M^*_N(\mathcal{S}) - M_N$. We see that this mass shift  is very close to the one generated by the quark core (the quark bag) alone. This is in line with the discussion given in section 2.1.1 of the review paper of Guichon et al (Ref. \cite{QMCreview}) concluding that the pionic corrections to the QMC model are moderate. There remain however some difficulties. In particular the value of the pionic contribution to the sigma term is rather low: $\sigma_N^{(\pi)}=18.5\,\mathrm{MeV}$, as well as the scalar field contribution $\sigma_N^{(s)}=22.5\,\mathrm{MeV}$. Hence  the total sigma term slightly exceeds $40\,\mathrm{MeV}$, below the accepted value of $50\,\mathrm{MeV}$. In addition the value of the response parameters are also lower that the one obtained in our papers using Bayesian analysis \cite{Rahul,Cham1,Cham2}. \\
 \begin{figure}
\includegraphics[width=10.5 cm]{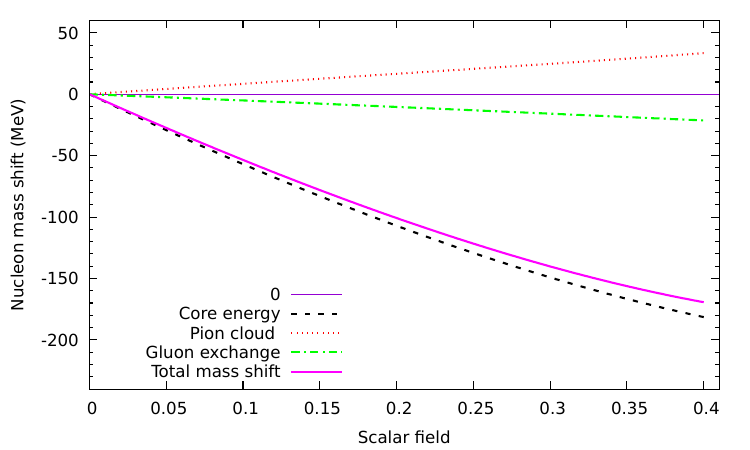}
\caption{The various contribution to the evolution of the nucleon mass, $\Delta M_N(\mathcal{S})=M^*_N(\mathcal{S}) - M_N$, versus $|s|/F_\pi$ : Dashed line: quark core contribution. Dotted line: pion cloud contribution. Dot-dashed line: gluon exchange contribution. Full line: total mass shift. \label{fig2}}
\end{figure}  
\section{Perspectives and conclusion}
An immediate perspective and source of improvement is to replace the factorized wave function by a translationally invariant nucleon wave function, together with a minimization with respect to $b$ at the level of the full mass to fully ensure  the mechanical stability of the nucleon. This will also provide another way to remove  the unwanted spurious CM contribution to the nucleon mass. We will briefly sketch this aspect in the following, postponing a more complete calculation to a future publication. \\

 For a nucleon state with CM momentum ${\bf P}_N$, a translationally invariant  normalized wave function can be taken with the form,
\begin{equation}
\Phi\left({\bf p}_1, {\bf p}_2, {\bf p}_3\right)=\frac{1}{V^{1/2}} N_{\bf{P}_N} \Phi\left({\bf p}_1\right)\,
\Phi\left({\bf p}_2\right)\,  
\Phi\left({\bf p}_3\right)\,(2\pi)^3\,\delta \left(\bf{P}_N - \bf{p}_1 -\bf{p}_2 - \bf{p}_3 \right),
\end{equation}    
where the normalization factor is :
$$
N_{\bf{P}_N}=\left(\frac{3}{4\pi b^2}\right)^{3/4}\,exp\left(\frac{P^2_N\, b^2}{6}\right).
$$
It is easy to check that the spatial Dirac wave function is manifestly translationally invariant:
\begin{equation}
\psi\left({\bf r}_1, {\bf r}_2, {\bf r}_3   \right)\sim
exp\left({i\bf P}_N\cdot {\bf R}_N\right)\,\int d^3X\,\psi({\bf r}_1-{\bf X})\,
\,\psi({\bf r}_2-{\bf X})\,\psi({\bf r}_3-{\bf X}),\label{CMWF}
\end{equation}
where ${\bf R}_N=\left({\bf r}_1+{\bf r}_2 +{\bf r}_3\right)/3$ is the CM position of the nucleon.\\
The replacement of the factorized wave function by the translationally invariant one has the following consequences:
\begin{itemize}
    \item 
    For the calculation of the expectation value of a momentum dependent operator such as the kinetic energy, one simply has to replace the $b^2$ parameter by $3b^2/2$. The consequence is to significantly reduce the contribution of high momentum quarks and then the kinetic energy itself.
    \item  To ensure translational invariance one can make make the replacement:
    $$\sum_j W_C({\bf x}_j={\bf r}_j-{\bf r}_0)\qquad\to\qquad \sum_j W_C({\bf r}_j-{\bf R}_N).$$
    For a two-quark system this amounts to replace $W_C(|{\bf r}_1-{\bf r}_0|)+W_C(|{\bf r}_2-{\bf r}_0|)$, by $2\,W_c(r/2)$, where $r=|{\bf r}_1-{\bf r}_2|$ is the relative distance between the two quarks; this is rather easy to understand since each individual quark is at the distance $r/2$ of the string junction located in the middle of the two quarks (see section 3 of Ref. \cite{Simonov99}). For the three-quark system the calculation of the potential energy is much more involved. Here just for orientation we assume that this is  equivalent to replace for each quark the single quark potential $W_c(r)$ by $W_C(2r/3)$, and replacing $b^2$ by $3b^2/2$. The accuracy of this approximation is probably not so crucial since the potential energy gives a much smaller contribution to the nucleon mass than the kinetic energy.  
    \item
    For what concerns the axial form factor, one can show that one has to make the replacement, 
    \begin{eqnarray}
&&v({\bf q})=\frac{5}{3}\,\int d^3r\,e^{i\bf{q}\cdot\bf{r}}\,\left(u^2(r)-\frac{1}{3}v^2(r)\right)\qquad\to\nonumber\\
&&\tilde{v}({\bf q})=\left(\frac{9}{8}\right)^{3/4}\,exp\left(\frac{q^2 b^2}{6}\right)\,\frac{5}{3}\,\int d^3r\,e^{i\bf{q}\cdot\bf{r}}\,\left(u(r)\tilde{u}(r)-\frac{1}{3}v(r)\tilde{v}(r)\right)\label{GACCM},
\end{eqnarray}    
where $\tilde{u}(r)$  and $\tilde{v}(r)$ correspond to the up and down Dirac wave function, but with $b^2$ replaced by $2\,b^2$. Similar results have been established in the past for the calculation of electromagnetic form factors including CM correction \cite{Tegen} . 
Keeping only the quark core contribution one finds a larger scalar coupling constant $g_S^{(core)}=7.38$, but a lower response parameter $C^{(core)}=0.245$. The pion cloud contribution remains small $C^{(core)}=-0.029$ but the full $C$ parameter remains nevertheless too small, of the order  of $C\sim0.2$. The next steps, that we reserve for a future publication, are first to perform the exact calculation of the potential energy with the translationally invariant wave function and second  to extract  the size parameter $b(\mathcal{S})$ from the three-quark wave function (Eq. \ref{CMWF}). The latter  point means that the minimization procedure  has to be done at the level of the total mass, namely $\partial M_N^*(\mathcal{S})/\partial b =0$. In that way the question of the mechanical stability (i.e., the pion pressure just compensating the Fermi pressure) will be exactly addressed.
\end{itemize}

 The general conclusion is that our QCD-based framework to generate the nucleonic response parameters to the nuclear scalar field gives the correct expected tendency. However both the generated scalar coupling constant, $g_S$, needed for sufficient nuclear binding, and the scalar susceptibility parameter, $C= F^2_\pi\, \kappa_{NS}/2\, M_N$, needed for the saturation mechanism, seem a little too small when compared with the values obtained in our recent Bayesian analysis \cite{Cham1,Cham2}. The same is also true for the in-medium Dirac mass at normal nuclear matter density, $M^*_N\sim 850 \,\mathrm{MeV}$, whereas our recent phenomenological studies \cite{Cham1,Cham2} favorize a lower value, 
 $M^*_N\sim 750-800 \,\mathrm{MeV}$. There are nevertheless some directions of improvement. The first one is to relax the simplest NJL approximation to generate chiral symmetry breaking.  However, as explained in section 3,  keeping the finite-range CSB interaction would imply the
use of bilocal meson fields resulting  in the probably long term development of a cumbersome formalism. A  more modest but realistic improvement would be to adapt the delocalized version of the NJL model presented in \cite{Chanfray2011}. Accordingly, the constituent quark mass would acquire an explicit momentum dependence which, together with a better treatment of the center of mass corrections discussed just above, would certainly improve the nucleon description and its mechanical stability. Indeed these mechanical properties of the nucleon open many possible connections between hadronic physics and compact stars as discussed in Ref. \cite{Lorce}. In particular one can envisage, as in Ref. \cite{Fuku}, a transition to a quark phase in the interior of neutrons stars based on a percolation criteria, which  needs the knowledge of the equation of state (energy density versus pressure) inside the quark core,  allowing to infer properties of dense quark matter from internal nucleon structure. \\
Finally from the point of view of pure nuclear physics phenomenology, there remains the possibility   that the  two identified origins of the scalar-isoscalar attraction (i.e., the "two-sigma" meson picture of Ref. \cite{Chanfray2024}) have to be reconsidered. In most of our previous papers, as in the QMC papers, only the first one, namely the $s=\sigma_W$ field associated with the radial fluctuation of the chiral condensate in the chirally broken vacuum, was considered.  Hence in  our recent paper \cite{Cham2}, where the NJL chiral effective potential was used, the Bayesian analysis yields $g_S\sim 10 $ and $C\sim 0.5$. Using smaller values of the parameters similar to those obtained in the present paper, $g_S\sim 6.5 $ and $C=0.32$, we found in Ref. \cite{Chanfray2024} that the contribution of a fictitious $\sigma'$ meson simulating two-pion exchange with $\Delta$'s in the intermediate state, as in Bonn-like OBE approaches, is needed to get sufficient binding. It also turns out that these two "sigma meson" contribute with similar strength the the NN scalar attraction in the nuclear medium. We also found that the associated three body-forces  contributing  to the saturation mechanism  and depicted in Fig. 1, are also similar in magnitude. The next task  is consequently  to improve the microscopic determination of the response parameters and more generally the evolution of the nucleon mass along  the lines developed above in parallel with Bayesian phenomenological calculations including the contribution of the correlation energy  which effectively contains the $\sigma'$ contribution. 

\appendix 
	
\section{Complement about the stochastic vacuum model and the field correlator method}
\label{sec:a1}
 For the interested reader we list some basic original papers  concerning the stochastic vacuum model and the field correlator method (FCM) used in this paper. The regularized area law, in the form of the surface-surface interaction mediated by the correlation function parameterized according to Eq. (14), was first considered in Ref. \cite{Mak}.
The vacuum correlation length in QCD was first discussed in the mid-1980s in the papers of the Pisa lattice group, such as \cite{Pisa}.  
The importance of the finiteness of the correlation length for confinement in QCD was first realized in Ref. \cite{Dosch}. 

Most of the analytic applications of the Stochastic Vacuum Model were devoted to the high-energy hadron-hadron scattering.  For reviews, see Refs. \cite{Dosch2,Donnach}.  These analytic studies were complemented by the lattice simulations of the two-point correlation function of gluonic field strengths, which were performed by the Pisa lattice group.  For reviews, see Refs. \cite{Meggio,Topics}.

\vspace{35pt}

%%%%%%%%%%%%%%%%%%%%%%%%%%%%%%%%%%%%%%%%%%
Author contributions.  Conceptualization, G.C., M.E., H.H., J.M. and M.M.; methodology, G.C., M.E., H.H, J.M. and M.M.; software, G.C.; validation, G.C., M.E, H.H., J.M. and M.M.; formal analysis, G.C..;  writing---original draft preparation, G.C.; writing---review and editing, G.C., M.E, H.H., J.M. and M.M. All authors have read and agreed to the published version of the manuscript.

\end{document}